\documentclass{aastex631}

\usepackage{xcolor}
\usepackage{hyperref}

\newcommand{\RMunit}{rad\,m\ensuremath{^{-2}}}

\newcommand{\Fig}{Fig.}

\newcommand{\Sect}{Section}

\newcommand{\lpt}{LPT\,J1555$-$5631 }
\newcommand{\lptp}{LPT\,J1555$-$5631}

\begin{document}

\title{Discovery of a 62-min long-period transient with near-orthogonal rotator geometry}

\author[0009-0006-5070-6329]{Scott D. Hyman}
\affiliation{Department of Engineering and Physics, Sweet Briar College, Sweet Briar, VA 24595, USA}

\author[0000-0002-5119-4808]{Natasha Hurley-Walker}
\affiliation{International Centre for Radio Astronomy Research, Curtin University, Kent Street, Bentley, WA, 6102, Australia}

\author[0009-0008-7604-003X]{Dale A. Frail}
\affiliation{National Radio Astronomy Observatory, P.O. Box O, Socorro, NM 87801, USA}

\author[0000-0003-3272-9237]{Emil Polisensky}
\affiliation{U.S.\ Naval Research Laboratory,  4555 Overlook Ave SW,  Washington,  DC 20375,  USA}

\author[0000-0001-7363-6489]{W. D. Cotton}
\affiliation{National Radio Astronomy Observatory, 520 Edgemont Road, Charlottesville, VA 22903, USA}

\correspondingauthor{Scott D. Hyman}
\email{shyman@sbc.edu}

\begin{abstract}
We report the serendipitous discovery of J155543.7$-$563102 (hereafter \lptp) with a period of 62.2 min in MeerKAT and ASKAP survey data. The source emitted bright (10-30 mJy), steep-spectrum ($\alpha\sim-1.5$), highly polarized radio pulses during a 15-hr active window in May 2022, with no other quiescent or pulsed emission detected before or after over $\sim5$ years. The mean light curve exhibits two alternating pulses per cycle. A main pulse (MP) and an interpulse (IP) separated by $\Delta\phi=165.4^\circ$ with similar amplitudes, but different polarization properties: the wider MP (3.9\%) is partially linearly and circularly polarized and shows a constant polarization position angle, while the narrow IP (1.1\%) is $\sim$100\% circularly polarized.  We classify \lpt as a near-orthogonal rotator, for which emission from both magnetic poles is visible during one rotation and the magnetic and rotation axes are almost perpendicular. The 14.6$^\circ$ deviation from an antipodal separation and the nearly constant polarization position angle suggest that the emission does not arise from the standard dipolar polar-cap geometry of ordinary pulsars, but instead originates within a broader, more complex magnetosphere. The observed properties of \lpt can be accommodated by extended emission regions, and favor coherent emission mechanisms such as the electron cyclotron maser instability, although they do not uniquely determine the progenitor or emission physics. If these inferences apply more broadly to the LPT population, they could account for why their beam widths are decoupled from rotation periods, the prevalence of flat polarization position angles, and may favor an elevated incidence of orthogonal rotators.
\end{abstract}

\section{Introduction} \label{sec:intro}
GCRT\,J1745$-$3009 stands as the enigmatic prototype of a newly emerging class of long-period radio transients \citep[LPT;][]{2026JHEAp..5200566R}. It was first discovered in September 2002 during a wide-field monitoring program of the Galactic center region with the Very Large Array \citep{2005Natur.434...50H}. Five bright (1 Jy) but short (10 min) bursts were detected with a period of 77.13 min. The pulsed emission was steep spectrum and highly polarized, and was likely coherent in origin. GCRT\,J1745$-$3009 had no counterpart at either X-ray or near-infrared wavelengths \citep{2005Natur.434...50H,2008ApJ...687..262K}. The lifetime of the radio source is constrained to be approximately 1.5 years.  This estimate is based on two short observations by the Giant Metrewave Radio Telescope which re-detected bursts from GCRT\,J1745$-$3009 on September 2003 and March 2004 \citep{2007ApJ...660L.121H,2010ApJ...712L...5R}. Despite on-going monitoring of the Galactic center \cite[e.g.,][]{2016ApJ...832...60P,2022MNRAS.516.5972W} and the analysis of archival data going back to 1989 \citep{2006ApJ...639..348H}, there has been no renewed activity from GCRT\,J1745$-$3009. Published upper limits for a quiescent radio source at this position are 0.4 mJy at 1.4 GHz and 15 mJy at 330 MHz \citep{2006ApJ...639..348H}.

While GCRT\,J1745$-$3009 remained an isolated anomaly for nearly two decades, the landscape of long-period radio transients has recently undergone a dramatic shift. Initiated by the discovery of GLEAM-X\,J1627$-$5235 \citep{2022Natur.601..526H}, the large increase in the number of LPTs today is being propelled by new synoptic imaging surveys \citep[e.g.,][]{2021PASA...38...54M,2022PASA...39...35H} and by powerful search and analysis tools \citep[e.g.,][]{2025PASA...42..129H,2026MNRAS.545f2008L}, designed to bridge the intermediate time domain phase space between fast pulsars and slow transients \citep{2026PASA...43....6M}. 

The newer LPTs \citep{2026JHEAp..5200566R, 2026arXiv260307857P,2026arXiv260604232R,2026arXiv260620067W} share some similarities with the original GCRT\,J1745$-$3009. Most if not all LPTS exhibit narrow, steep-spectrum polarized pulses with periods from minutes to hours \citep{2026JHEAp..5200566R}. Their active lifetimes however vary from less than a month to years, even decades for a few \citep{2023Natur.619..487H,2024ApJ...976L..21H}. Whether LPTs can reactivate after entering dormancy remains unclear. GCRT\,J1745$-$3009 exhibited multiple bursts over $\sim$1.5 years, but has not been detected despite continued monitoring. No LPT has been observed to transition from an active to a dormant state and back again, and no LPT has been detected in a quiescent, non-pulsed state. While some sources lack multi-wavelength counterparts \citep[e.g.,][]{2022Natur.601..526H}, others show optical evidence of white dwarf (WD) and M dwarf binary companions \citep{2025A&A...695L...8R,2025NatAs...9..672D,2026arXiv260418688R}. In other LPTs a link to neutron stars (NS) is suggested by pulsar-like behavior including pulse/interpulse, mode switching, glitches, and quasi-periodic sub-pulse structure \citep[e.g.,][]{2024NatAs...8.1159C,2024NatAs...8..230K,2025NatAs...9..393L,2025ApJ...988L..29D}.

Long-period transients (LPTs) are currently defined phenomenologically rather than physically and thus leave many questions unanswered. Are the progenitors of LPTs a single object class with diverse observational states, or do they represent distinct populations of WDs and NSs, that simply overlap in the radio time-domain? Can their pulsed emission properties be explained by a pulsar-like coherent curvature radiation, or electron cyclotron maser emission, or some other alternate plasma mechanism? Furthermore, what physical triggers dictate the onset and duration of their intense active phases before many lapse back into dormancy? And can estimates of the rotation measure (RM) and dispersion measure (DM) constrain the immediate magneto-ionic environment around LPTs?

Progress in addressing these questions requires that we increase the sparse sample of LPTs \citep{2026MNRAS.545f2008L}. In this paper we report on the serendipitous discovery of LPT\,J155543.7$-$563102 (hereafter \lptp) with a 62-minute period. A study of its pulse morphology establishes \lpt as a near-orthogonal rotator with a large-scale dipolar field that is significantly distorted from a pure dipole. The broad, non-canonical polar-cap beam geometry and extreme circular polarization place constraints on the emission mechanism and magnetospheric geometry that we use to draw comparisons with the broader LPT population. In \S\ref{sec:obs} and \S\ref{sec:multi} we describe the observations and data reduction. In \S\ref{sec:results} we summarize the key observational results and then in \S\ref{sec:discuss} we look to interpret the properties of \lptp, compare it to the existing sample of LPTs, and suggest avenues for future progress.

\section{Radio Observations}\label{sec:obs}

\subsection{MeerKAT and ASKAP Discovery}\label{sec:discover}

\begin{figure}
\centering
\includegraphics[width=18cm]{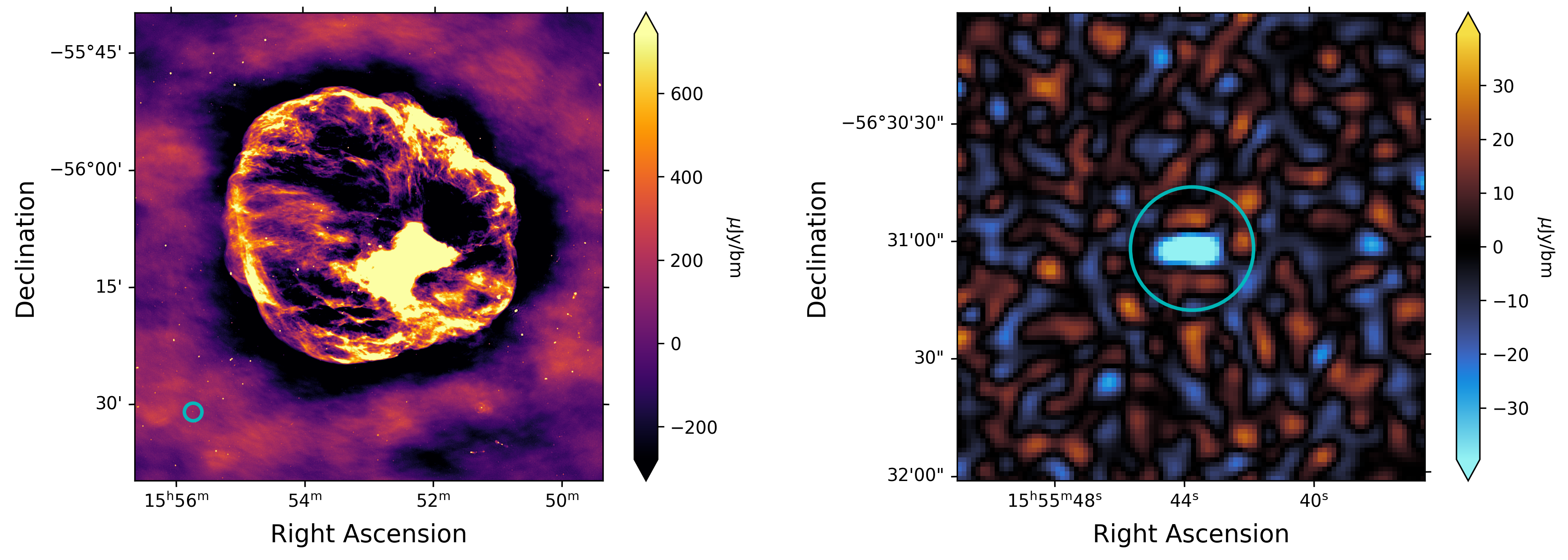}
\caption{MeerKAT 1.3 GHz discovery images of \lpt {\it Left}: Stokes I image of the full field centered on the SNR G\,326.3$-$1.8. The blue circle marks the position of the LPT. {\it Right}: Stokes V cutout (2' $\times$ 2') centered on the LPT, again circled. The source appears strongly circularly polarized (negative V); its elongated morphology is an imaging artifact possibly due to strong intra-observation variability and/or baseline-dependent time averaging, which smears the emission along the instantaneous point-spread function.}\label{fig:g326.31.8}
\end{figure}

\lpt was discovered as part of a search for steep spectrum, circularly polarized point sources following the methodology outlined in \citet{2024ApJ...975...34F}. The original data came from a public archive\footnote{\url{https://archive-gw-1.kat.ac.za/public/repository/10.48479/nz0n-p845/index.html}} \citep{https://doi.org/10.48479/nz0n-p845} of the South African Radio Astronomy Observatory (SARAO), taken by the 64-element MeerKAT Radio Telescope Array \citep{2016mks..confE...1J}.  \citet{2024ApJS..270...21C} observed 36 high-latitude supernova remnants (SNR) in full Stokes at L-band (856-1712 MHz). The composite remnant G\,326.3$-$1.8 (MSH\,15$-$56) was observed as part of the SNR survey over an 8 hour period on 2022 May 14, with a total of 2 hours (4 x 30 minutes) of time-on-source. On the initial Stokes I and V images \lpt appeared as a 0.49 mJy, polarized ($\vert{\rm{V/I}}\vert=34\%$), steep spectrum ($\alpha=-1.1$) source with a distorted point-spread function - suggestive of short-term variability. The position in Galactic coordinates is $(l,b)$=(326.36$^\circ$, $-2.27^\circ$), or 0.52$^\circ$ from the center of the SNR. As there is a known pulsar powering G\,326.3$-$1.8 \citep{2017ApJ...851..128T}, it is unlikely that there is a physical association between \lpt and the SNR. See Fig.\,\ref{fig:g326.31.8} for full-field Stokes I and V images.

By good fortune, \lpt was observed contemporaneously by the 36-element Australian Square Kilometre Array Pathfinder (ASKAP) as part of the Evolutionary Map of the Universe \citep[EMU;][]{2021PASA...38...46N} 144 MHz-wide pilot survey at 1.367 GHz. We downloaded archival data for the 14 May 2022 epoch as well as all available catalogs from the CSIRO ASKAP Science Data Archive\footnote{\url{https://data.csiro.au/domain/casdaObservation}}. Out of 59 epoch catalogs, \lpt is found only in the 14 May 2022 epoch. The 1.367 GHz cataloged flux density is 0.47 mJy. Images were made similar to Fig. \ref{fig:g326.31.8} using the beam pointed nearest to the source. 

\begin{figure}
\includegraphics[width=18cm]{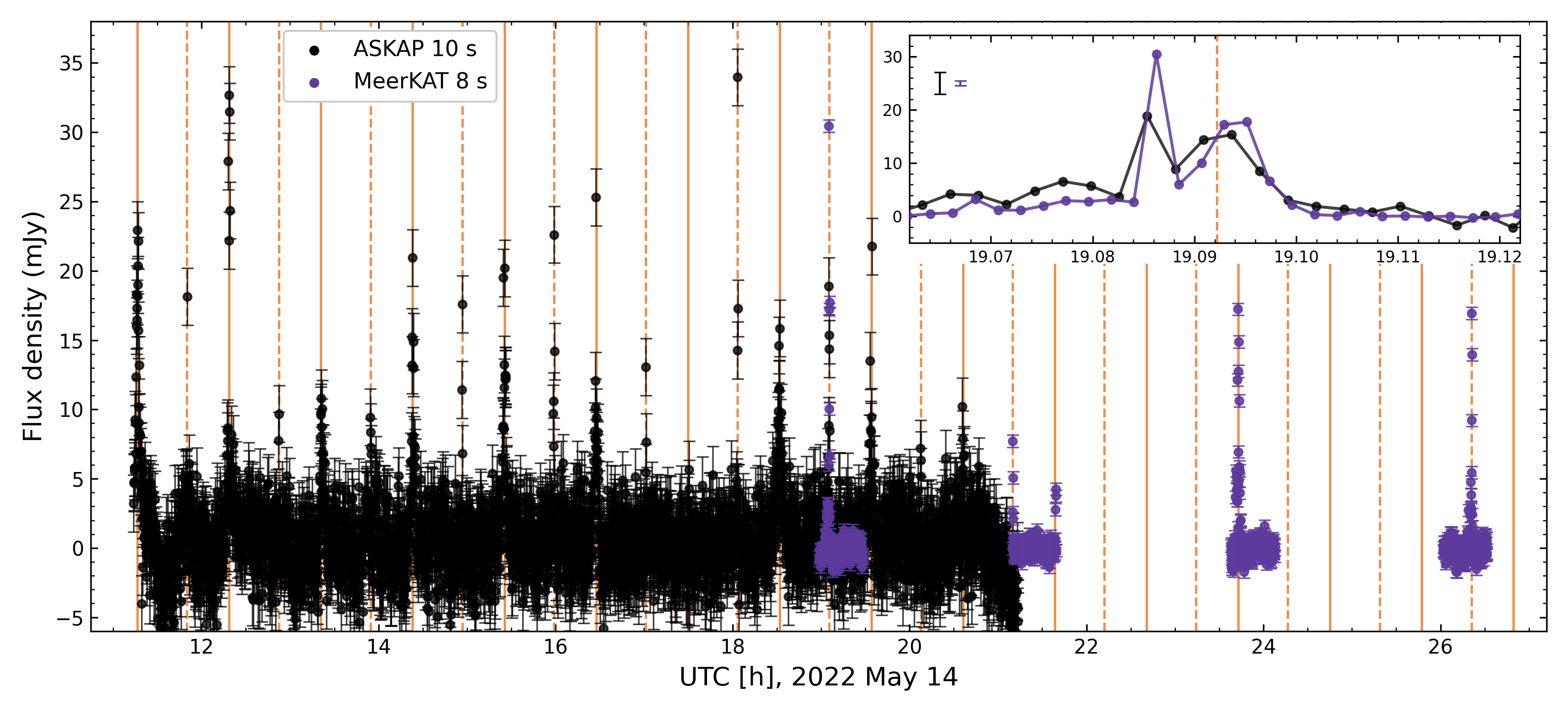}
\centering
\caption{A composite light curve of \lpt made from the 1.36 GHz ASKAP (black) and 1.28 GHz MeerKAT (purple) {\it uv} datasets, at full 10 s (ASKAP) and 8 s (MeerKAT) time resolution. Solid (dashed) vertical lines mark the main pulse (interpulse) peaks from the timing ephemeris described in \S\ref{sec:analysis}. The inset shows the simultaneous ASKAP and MeerKAT interpulse (see \S\ref{sec:analysis}); representative error bars are shown at the top left of the inset.}\label{fig:lightcurves_1030}
\end{figure}

To make an initial Stokes I light curve we followed a process first outlined by \citet{1986syim.conf..123C} in which a time-variable point source is separated out from the constant (and extended) emission in the field by subtracting CLEAN components of the latter from the {\it uv} plane. A vector average of these model-subtracted visibilities was made at the position of \lpt, initially with 30-sec averaging. The ASKAP and MeerKAT datasets were combined after correcting for the primary beam of each telescope, resulting in a clear periodic burst-like signature. Subsequent analysis at the full 10-sec (ASKAP) and 8-sec (MeerKAT) time resolutions (see \S\ref{sec:analysis}) produced the light curve shown in Fig. \ref{fig:lightcurves_1030}. 

Inspection of the light curve revealed that consecutive bursts exhibit alternating pulse widths, alternating polarization properties, and systematically unequal time separations. This emission behavior is similar to ASKAP J183950.5$-$075635.0 which has been suggested to originate from interleaved main pulses and interpulses originating from opposite magnetic poles \citep{2025NatAs...9..393L}. See \S\ref{sec:analysis} for more details.

The best-fit position for \lpt was determined by imaging only the MeerKAT and ASKAP data during outbursts, when the source was brightest. The average of the MeerKAT and ASKAP positions is (J2000) R.A=15:55:43.67$\pm{0.12}$s, Dec.=$-$56:31:02.6$\pm{1}^{\prime\prime}$, with an offset of only 0.21$^{\prime\prime}$ between the two. The uncertainties are based on a comparison of the position offsets for 156 and 166 radio sources in common to the MeerKAT image and to ASKAP images from the RACs-mid and RACS-high \citep{2024PASA...41....3D, 2025PASA...42...38D}, respectively. The standard deviations in position for bright sources was of $\pm{1}^{\prime\prime}$ which we adopt as a position error, and note that it includes $\sim{0.1}^{\prime\prime}$ due to uncorrected systematics in these early MeerKAT and ASKAP observations \citep[e.g,][]{2024ApJS..270...21C}. Our quoted position uncertainty is consistent with other LPTs found by ASKAP \citep[e.g.,][]{2025NatAs...9..393L}.

\subsection{Refined Data Analysis}\label{sec:analysis}

\begin{figure*}
    \centering
  \includegraphics[width=\linewidth]{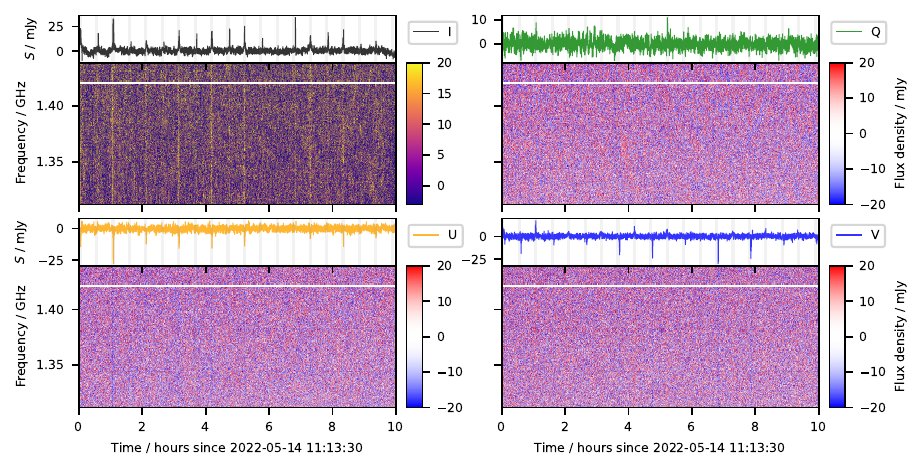}
    \caption{ASKAP SB ID 40625 dynamic spectra (panels) and frequency-averaged light curves (above each panel) for Stokes I, Q, U and V. Pulse arrival times derived from the ephemeris are indicated with vertical grey lines (see \S\ref{sec:analysis}). The RMS noise of the light curves is 2\,mJy\,beam$^{-1}$.}
    \label{fig:askap_dynspec}
\end{figure*}

Following the initial detection of \lpt (\S\ref{sec:discover}), we undertook a more detailed analysis of the 2022 ASKAP and MeerKAT datasets. LPT\,J1555$-$5631 lies in the main fields of both beam 9 and 15 of the EMU pilot observation SB ID 40625. We downloaded the calibrated measurement sets from CASDA and used the \textsc{FixMS} software\footnote{\href{https://github.com/AlecThomson/FixMS}{https://github.com/AlecThomson/FixMS}}  \citep{Thomson2026AlecThomson} to phase rotate each beam to its peak primary beam sensitivity direction, and correct the instrumental Stokes parameters to the IAU standard. We performed Stokes~I imaging of each beam using \textsc{WSclean} \citep{2014MNRAS.444..606O}, with a $10\times10$ pixel mask at the location of the source. This populates a new table in the measurement set with the Fourier transform of the clean components, referred to as the ``MODEL'' column. We then subtracted the now-populated MODEL column of each measurement set from the DATA column (the original calibrated visibilities). This has the effect of subtracting the non-target ``field'' sources elsewhere in each primary beam from the visibility data, leaving only the target LPT. We then phase-rotated to the source location and averaged the visibilities over all baselines, preserving the time, frequency, and polarization dimensions, to form dynamic spectra. We applied primary beam and leakage corrections using the \textsc{ASKAPSoft} routine \textsc{linmos} \citep{https://doi.org/10.25919/5cca3787a6353}. We observed that the polarization behaviour of the dynamic spectra of each beam was identical, and therefore averaged them together, using a weighting inversely proportional to their required primary beam correction (i.e., upweighting beam~15 slightly, where the source is closer to the center). \Fig~\ref{fig:askap_dynspec} shows the Stokes I, Q, U, and V dynamic spectra and frequency-averaged light curves. We note that the pulses appear alternately linearly- and circularly-polarized, with notable negative flux density alternating between U and V, with Q only producing two notable, weaker peaks.

\begin{figure*}
    \centering
  \includegraphics[width=\linewidth]{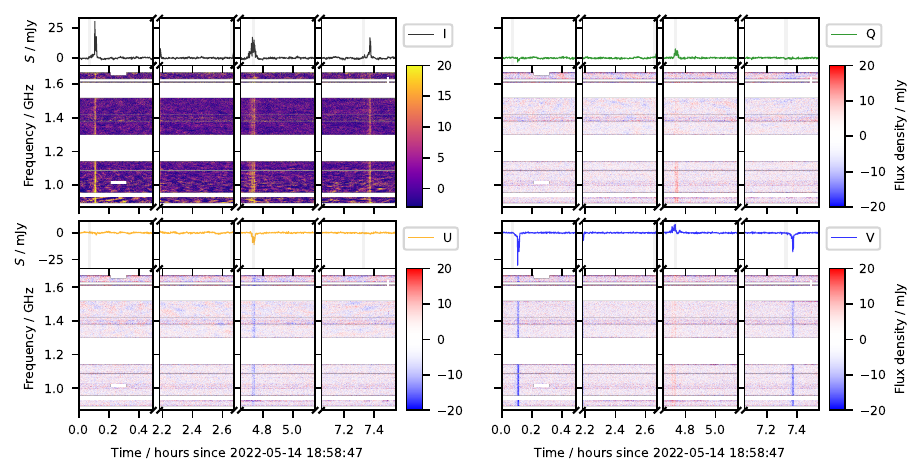}
    \caption{MeerKAT CBID 1652551867 dynamic spectra (panels) and frequency-averaged light curves (above each panel) for Stokes I, Q, U and V. Pulse arrival times derived from the ephemeris are indicated with vertical grey lines (see \S\ref{sec:analysis}). The RMS noise of the light curves is 0.44\,mJy\,beam$^{-1}$.}
    \label{fig:meerkat_dynspec}
\end{figure*}

The MeerKAT data were downloaded from the SARAO archive in ``uvfits'' format applying online flagging but not calibration. The data were then calibrated using \textsc{Obit}, running a standard calibration script. The initial X-Y phase and delay calibration was based on the noise diode calibration in the DelayCal performed before the observing session.  Then the unpolarized calibrators J0408$-$6545 and J1939$-$6342 were used to calibrate the group delay and bandpass for the two parallel--hand (XX,YY) systems.  The gain calibration used J1424$-$4913 and J1726$-$5529 on Stokes~I for astrometric and photometic calibration.  The amplitude scale is set by the standard model for J1939$-$6342. Instrumental polarization (AKA ``leakage'') was determined from the unpolarized calibrators J0408$-$6545 and J1939$-$6342.  Since the observation had no known polarized calibrator, the residual X-Y phase and delay correction used the average of multiple, similarly calibrated, datasets which did contain a known polarized calibrator. We obtained dynamic spectra (\Fig~\ref{fig:meerkat_dynspec}) by following a similar field-subtraction process to the ASKAP data, with the addition of applying a parallactic angle correction when determining Stokes~Q and U. A strong negative circularly polarized pulse is detected in the first and fourth 30-min scans, the former overlapping the EMU pulse at 19:05. A strongly linearly polarized pulse is detected in the third scan.

\begin{figure}
\includegraphics[width=\linewidth]{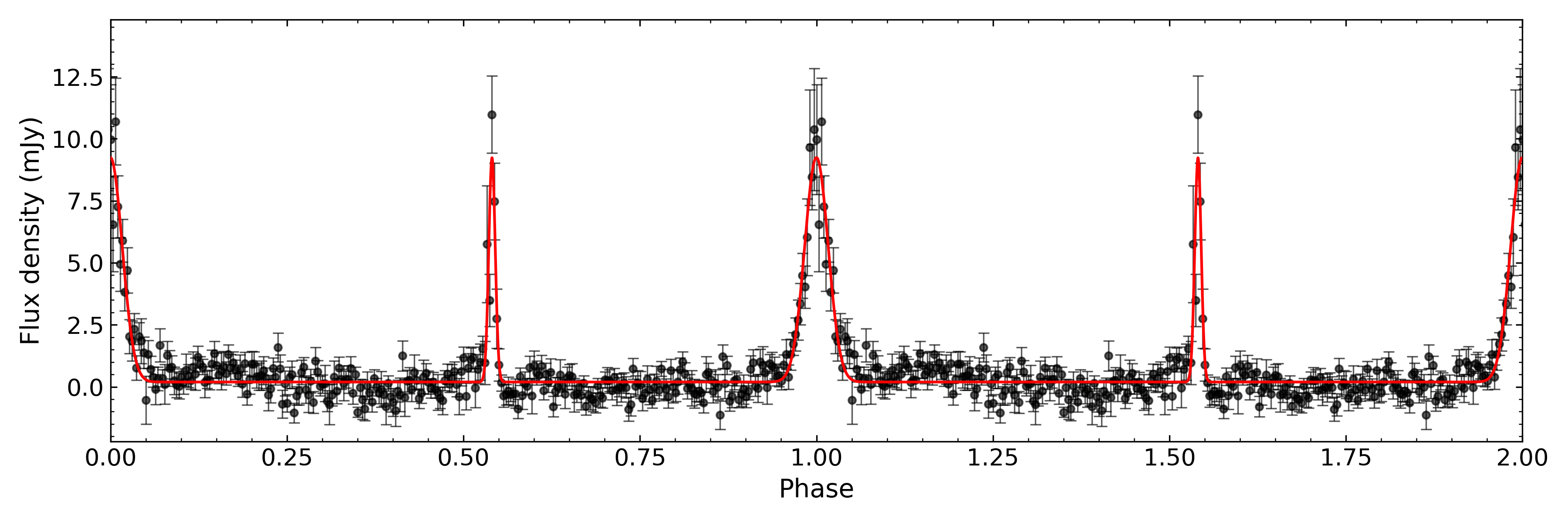}
\centering
\caption{Folded pulse profile of LPT\,J1555$-$5631 at the refined period of $62.2$~min, constructed from the combined ASKAP and MeerKAT datasets. Each data point represents the mean flux density in a phase bin of width $1/300$ of the period. The red curve shows the best-fit model comprising two Gaussian components on a periodic baseline, corresponding to the main pulse (phase $\sim$0.0) and interpulse (phase $\sim$0.54).}\label{fig:folded_profiles}
\end{figure}

Using the combined 10-s ASKAP and 8-s MeerKAT datasets, we calculated a Lomb-Scargle periodogram \citep{2022ApJ...935..167A} to obtain an initial period estimate. The light curve was then folded at this period and the folded profile was constructed by computing the mean flux density in each of 300 uniformly spaced phase bins. The resulting profile exhibits two clearly separated peaks corresponding to a wide pulse and narrower second pulse (Fig.~\ref{fig:folded_profiles}). We fitted the profile with a model consisting of two Gaussian components on a periodic domain plus a constant baseline (seven free parameters in total) using nonlinear least squares \citep{2020NaMet..17..261V}. Initial centroid positions were determined automatically from the two tallest peaks in the baseline-subtracted profile. The component with the wider extent is designated the main pulse as the amplitudes of the two components are statistically indistinguishable, making width the better discriminant. Fitted values are found in Table \ref{table:properties}.

The Gaussian centroids provided an initial estimate of the pulse phase offset and the absolute phase of each pulse within each cycle. Using this phase information to predict the expected arrival time of each individual pulse, we then identified the brightest data point above a signal-to-noise threshold of 5-$\sigma$ within a window of $\pm8\%$ of the period centered on each predicted pulse time. A straight-line fit $T(n) = T_0 + nP$ to these observed peak times by ordinary least squares yielded a refined period and a precise reference epoch $T_0$, with uncertainties estimated from the RMS scatter of the timing residuals. Finally, the light curve was re-folded at the refined period and the two-Gaussian model was re-fitted to obtain the final profile parameters.

We derive a best-fit period of $0.04320 \pm 0.00001$~d ($62.20 \pm 0.02$~min) and a MJD reference epoch $T_0 = 59713.4266 \pm 0.0001$~d. A second pulse arrives $0.540 \pm 0.001$ cycles after the main pulse, offset from strict half-period by $0.040$ in phase ($2.5$~min). Pulse widths at 50\% and 10\% of peak amplitude ($W_{50}$ and $W_{90}$) were computed analytically from the fitted Gaussian widths. The main pulse is wider ($W_{50} = 2.44 \pm 0.09$~min, $W_{90} = 4.44 \pm 0.16$~min) than the second pulse ($W_{50} = 0.66 \pm 0.04$~min, $W_{90} = 1.21 \pm 0.08$~min), consistent with the alternating pulse widths that originally suggested a main pulse/interpulse identification. The results of the ephemeris and Gaussian profile fitting are summarized in Table~\ref{table:properties}.

\begin{figure}
    \centering
    \includegraphics[width=\linewidth]{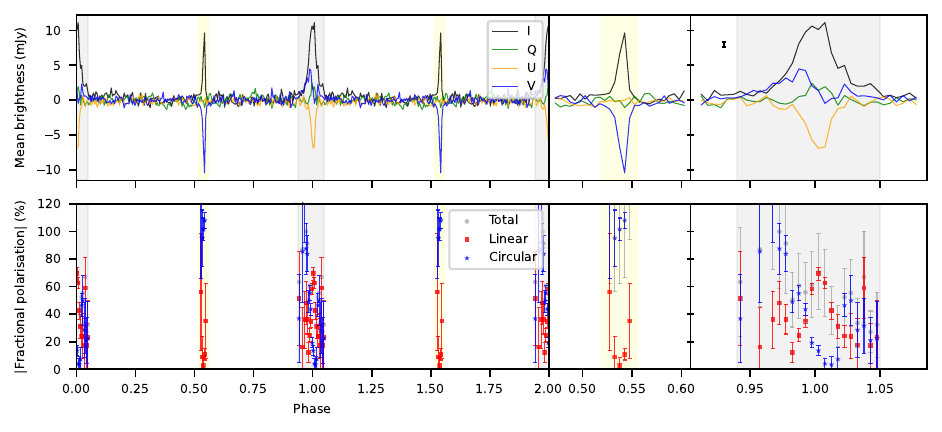}
    \caption{EMU SB ID 40625 light curve data folded on the ephemeris listed in Table \ref{table:properties}, across 200~phase~bins. Approximately 9 periods are folded, leading to a factor of 3 decrease in RMS noise (to $\sim0.7$\,mJy\,beam$^{-1}$, shown with a representative error bar in the top left of the top right panel). The top panels show the celestial Stokes parameters while the lower panels show derived polarization fractions. Only the shaded regions have sufficient signal-to-noise to display this information. The sub-panels in the middle and right show zooms of the full phase information presented on the left. The grey shaded area covers the main pulse and was analysed with RM synthesis in \Sect~\ref{sec:analysis}, finding a polarization position angle of 142$^\circ$, with no significant variation over the folded pulse.}
    \label{fig:askap_fold}
\end{figure}

The folded ASKAP polarization profile reveals a clear difference in polarization between the two alternating pulses (Fig.~\ref{fig:askap_fold}). One pulse is $\sim$100\% negatively circularly polarized, while the other pulse exhibits a peak of $\sim$90\% positive circular polarization at the start of the pulse and $\sim$75\% linear polarization midway through. Folding the data stacks 9 periods, reducing the noise by a factor of $\sim$3, and reveals a small positive Stokes Q component in contrast to the negative Stokes U. We did not attempt to fold-in the three MeerKAT pulses due to the very different bandwidths of the two telescopes.

The polarization angle of linearly-polarized sources is Faraday rotated by the parallel component along the line-of-sight of magnetic fields $B_\parallel$ in the ionized (with $n_e=$ electron density) interstellar medium, proportionate to $\int{n_eB_\parallel}$. This effect is stronger as a function of $\lambda^2$ and thus a Stokes spectrum can be used to determine the rotation measure (RM), as well as the polarization angle of the linear polarization.

\begin{figure}
    \centering
   \includegraphics[width=0.6\linewidth]{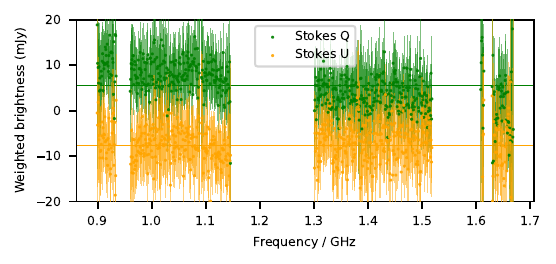}
   \includegraphics[width=0.35\linewidth]{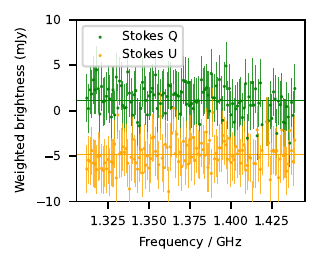}
    \caption{The Stokes Q and U spectra as a function of frequency. Left: the main pulse detected by MeerKAT. Right: the folded main pulse data as measured by ASKAP (right-most grey-highlighted region in \Fig~\ref{fig:askap_fold}). To maximise S/N, these are weighted by the Stokes~I brightness of the frequency-averaged pulse profile in each phase bin, then averaged over phase without frequency-averaging. The horizontal lines show the (unweighted) average over all frequencies of each Stokes parameter, to make the relative brightnesses of each parameter more clear. Very little rotation of Q and U is evident across the band, leading to a low RM estimates derived in \Sect~\ref{sec:analysis}.}
    \label{fig:weighted_askap_stokes}
\end{figure}

For the MeerKAT data, only the main pulse showed sufficient signal in Q and U to perform this measurement. For the ASKAP data, with its lower signal-to-noise in the individual pulses, we performed the analysis on the weighted stack of the broader pulse. We selected the phase bins containing this pulse (right-most grey-highlighted region in \Fig~\ref{fig:askap_fold}) from the ASKAP dynamic spectrum, and weighted them by the Stokes~I brightness of the light curve at each location, setting any weights with a negative Stokes~I flux density to zero. For each dataset, we performed a weighted average over the phase dimension and obtained the spectra shown in \Fig~\ref{fig:weighted_askap_stokes}.

We used RM-Tools \citep{2026ApJS..283...28V} to perform RM synthesis. The RM is more effectively measured by MeerKAT due to its far wider bandwidth as $4.5\pm0.7$\,\RMunit{}; ASKAP's measurement is slightly higher but far more poorly-constrained: $6\pm10$\,\RMunit{}. The relative proportions of Q and U yield a strong estimate of the polarization angle of 142$\pm1^\circ$ (ASKAP) and 156$\pm1^\circ$ (MeerKAT). We also attempted to bin the data in phase more finely, to determine whether the polarization angle changed over the pulse, but found no statistically-significant difference in the polarization angle between phase bins. We also attempted to determine a polarization angle from the ASKAP measurements of each pulse separately, but recovered the same value, with variations entirely attributable to the higher noise level. Given this, we expect that the $\sim$14$^\circ$ difference in measured polarization angle between the two instruments is due to unmodelled MeerKAT primary beam errors for this source that lies well off the boresight.

With these data it is not possible to provide a strong distance constraint to \lptp. We attempted to derive a dispersion measure by looking at individual bright pulses and sub-structure within pulses (\Fig~\ref{fig:acf}). Unfortunately we were not able to measure a dispersive delay for such broad pulses at these frequencies and bandwidths, and the resulting upper limit on DM is not useful for a distance estimate. The RM can in principle provide a secondary distance estimate, as there is a rough scaling of DM with RM but with large scatter \citep[e.g.,][]{2026JHEAp..5200566R}. From the ATNF Pulsar Catalogue v2.8.0 \citep{2005AJ....129.1993M} we identified four known pulsars within 1 to 2.5 degrees of \lpt which have RMs of comparable magnitude (10--25\,\RMunit{}). The DM-distances for these four pulsar ranges from 3.7 to 6.5\,kpc, with a mean of 5.4\,kpc. While the RM may vary smoothly on angular scales of degrees, there are significant RM variations at the scale of several arcminutes, especially at low Galactic latitudes \citep{2026arXiv260516917T}. Thus this distance should be used with caution in deriving any physical quantities. 

\subsection{Individual Pulse Analysis}\label{sec:single}

\begin{figure}
    \centering
   \includegraphics[width=0.3\linewidth]{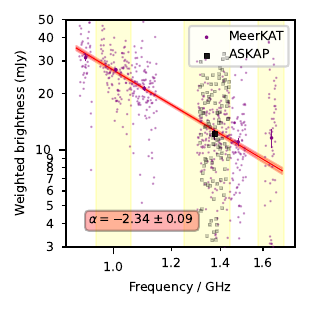}\includegraphics[width=0.3\linewidth]{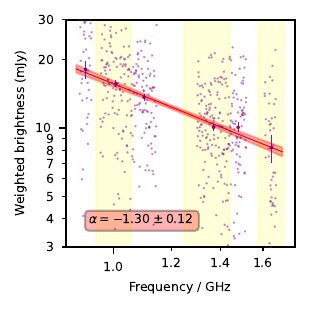}\includegraphics[width=0.3\linewidth]{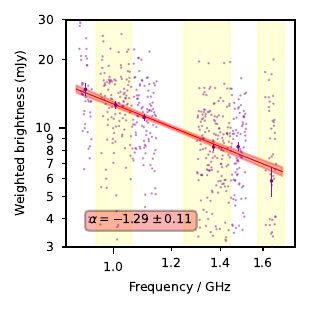}
    \caption{The flux density as a function of frequency for the three pulses detected by MeerKAT. Left panel: the single interpulse detected at 19:05 UTC on 2014 May 14 as measured simultaneously by both MeerKAT (purple circles) and ASKAP (black square). Middle panel: the main pulse detected by MeerKAT (Pulse 13 in \Fig~\ref{fig:acf}). Right panel: the second interpulse detected by MeerKAT at 02:21 UTC on 2015 May 15. The transparent points show individual (low signal-to-noise) channel measurements, and the fully-shaded points show the binned data. The alternating yellow shading indicates the binning selections for the MeerKAT data. A power-law fit to the data (with 1-quartile error regions shaded) is shown in red, with the resulting fitted spectral indices $\alpha$ printed on each panel.}
    \label{fig:joint_spec}
\end{figure}

The interpulse at 19:05 UTC is jointly detected by both ASKAP and MeerKAT (see \Fig~\ref{fig:lightcurves_1030}), enabling a test of the flux density calibration between the two telescopes, and a measurement of the spectrum. We selected the same 45~seconds of the Stokes~I light curves during which the pulse became bright ($>10$\,mJy) and averaged the dynamic spectra for that section, weighting by the brightness of the pulse, and preserving the frequency dimension. We then binned the MeerKAT data into six subbands to boost the signal-to-noise, and averaged all of the ASKAP data into a single measurement. We found that the telescopes agreed to within measurement uncertainties (\Fig~\ref{fig:joint_spec}). We fitted a power-law spectrum to the concatenated, binned datasets, finding that the spectral index $\alpha=-2.34\pm0.09$. A similar process was followed for the MeerKAT main pulse at 23:42 and interpulse at 02:21 on 15 May 2022, resulting in $\alpha=-1.30\pm0.12$ and -1.29$\pm$0.11, respectively. Spectral indices were also determined from the Stokes V data for the MeerKAT interpulses yielding consistent results.

\begin{figure*}
    \centering
    \includegraphics[width=\linewidth]{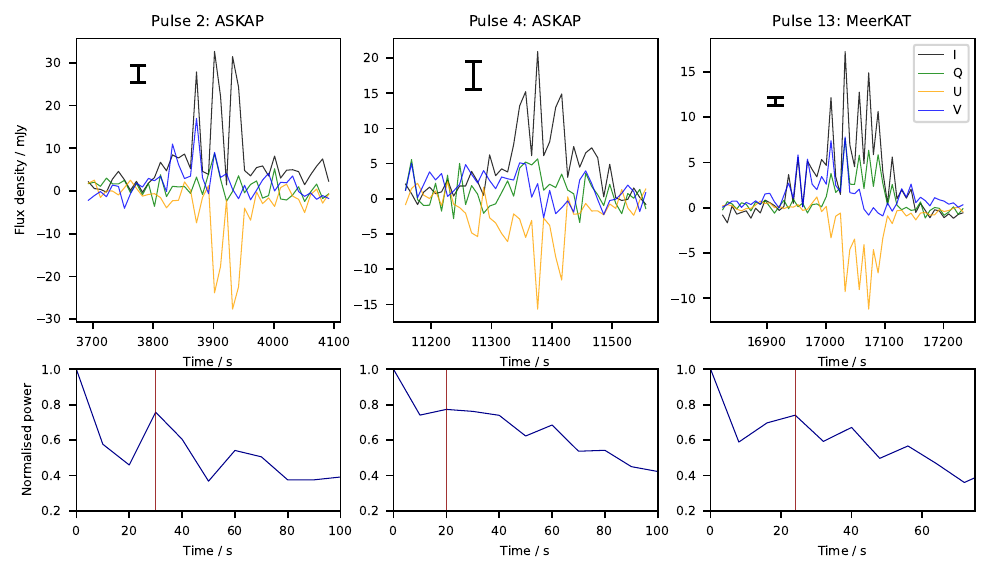}
    \caption{Autocorrelation analysis for the three main pulses that show quasi-periodic substructure. The top panels show the polarized brightnesses of the source as a function of time in seconds since the start of their respective observations on target. Representative ($\pm$1-$\sigma$, derived off-pulse) error bars are shown at the top left of each panel. The lower panels show the autocorrelation functions for each dataset, with the highest peak (excepting lag$=0$) highlighted with a red vertical line.\label{fig:acf}}
\end{figure*}

Several of the main pulses (namely 2, 4, and 13) displayed interesting quasi-periodic sub-structure (QPSS; Fig. \ref{fig:acf}). To quantitatively estimate the timescale of the periodicity, we performed a single-pulse autocorrelation analysis similar to that of \citep{2024ApJ...976L..21H}, wherein the pulses are cross-correlated with increasing lag separation on the time sampling intervals (10\,s for ASKAP, 8\,s for MeerKAT). The results are shown in \Fig~\ref{fig:acf}, demonstrating 30-, 20-, and 24-s periods within pulses 2, 4, and 13, respectively. 

All the interpulses are nearly 100\% circularly polarized. The peaks of the three main pulses shown in \Fig~\ref{fig:acf} and the other ASKAP ones have linear polarization fractions with mean and standard deviation of 0.80$\pm$0.18. The figure also shows that the Stokes $V$ flux of the main MeerKAT pulse 13 develops prior to the Stokes U flux as seen in the folded ASKAP main pulse of \Fig~\ref{fig:askap_fold}.

Finally, we note that faint inter- and main pulses are detected at the beginning and end of the second $\sim$30-min MeerKAT scan (see \Fig~\ref{fig:lightcurves_1030}), as predicted by the ephemeris. We did not analyze them further since they are cut off at the edges of the scan.

\subsection{Additional Radio Archival Datasets}\label{sec:archive}

In order to constrain the active lifetime of the radio source we downloaded relevant MeerKAT datasets and ASKAP images from the archives covering from 2019 to 2026, listed in Table \ref{tab:radio}. The largest number of epochs come from the ASKAP archive, totaling about 60 epochs. The suitability of epochs varies. While all are observed within the spectral window that defines L band, some are short integrations, some are pointed far from \lpt, and in all but one (2022 June 25) our ephemeris is too inaccurate to determine whether a pulse should lie within the observing window.

\subsubsection{ASKAP}\label{sec:askap_singlepulse}

Following the procedure described in \Sect~\ref{sec:analysis}, we downloaded, imaged, and produced dynamic spectra from all relevant PAF beam visibility datasets from CASDA. Where an individual observation comprised multiple beams, data from these beams were primary-beam-corrected and averaged together (weighted by the primary beam correction, to upweight measurements made closer to the primary beam center). After channel-based RFI flagging, Stokes~I light curves were formed by averaging along the frequency axis.  The resulting light curves are shown in \Fig~\ref{fig:archive_search}, folded on the ephemeris derived in \S\ref{sec:analysis}. Due to the short time baseline over which the ephemeris could be calculated, the uncertainty swiftly propagates, rendering almost every observation potentially able to contain a pulse (light green shading). Nevertheless, no pulses are seen, likely indicating that the source had a very short active window. The lower limit on this activity window of 15 hrs comes from the total duration of the joint MeerKAT and ASKAP May 2022 observations. An upper limit on the activity lifetime of 10 months comes from observations taken on either side of this epoch (2021 Nov. 06 and 2022 Sept. 07) with sufficient sensitivity and phase coverage to detect pulses (Table \ref{tab:radio} and Fig. \ref{fig:archive_search}). 

To determine if the individual archival 887 MHz VAST images would have detected \lpt, we estimated the flux for a typical pulse occurring in a 12 min observation. The 900 MHz flux in our 2 hr MeerKAT average spectrum is 0.8 mJy. For the VAST epochs, though, the average flux would be 2 mJy after scaling by the ratio of the 31 min period to the 12 min observation. The typical RMS noise levels of the VAST images is 0.3\,mJy\,beam$^{-1}$, which would therefore easily have detected a typical pulse. The bottom panel of \Fig~\ref{fig:archive_search} plots the brightest detections and the upper limits of the archival observations at the 10-s ASKAP native time resolution in comparison. In addition to the detection in May 2022, there were 56 other ASKAP observations toward \lpt with no evidence for any earlier or renewed pulsed activity. The 3-$\sigma$ upper limit for any pulses from \lpt for the year-long interval after May 2022 is typically 4 mJy (\Fig~\ref{fig:archive_search}).

\subsubsection{MWA}

The Galactic Plane Monitor (Hurley-Walker et al. accepted) is a 200-MHz 72-epoch survey of the Galactic plane undertaken with the Murchison Widefield Array \citep[MWA;][]{2013PASA...30....7T,2018PASA...35...33W}.
Four 30-minute integrations were taken on a pointing containing \lpt across 2022 June 02 to 2022 June 23 inclusive. No detection was made on any timescale to which the survey is sensitive (4\,s to 30\,min) with RMS noise levels of 100\,mJy\,beam$^{-1}$ to 5\,mJy\,beam$^{-1}$. Several LPTs have steep spectral turnovers that have rendered them invisible at low frequencies \citep{2025NatAs...9..393L,2025Natur.642..583W}, but due to the lack of simultaneity in the observing, it is unclear whether this non-detection yields a spectral or activity constraint.

\subsubsection{MeerKAT}

In addition to the 2022 May 14 discovery epoch, the MeerKAT archives contain 11 L-band 10-min observations pointed within 0.7 deg (the 33\% point of the L-band primary beam) of \lptp from the 2021 MPIfR Galactic Plane Survey setup program. The nearest of these to the discovery epoch was made on 2021 December 22. None detect \lptp. Two 15-min observations from the ThunderKAT program were made on 2019 November 30 and December 7. The RMS noise level of each is 80\,$\mu$Jy\,beam$^{-1}$.

\subsubsection{Quiescent Counterpart}\label{sec:quiet}

In addition to searching for new radio pulses, we used the archival data to put 3-$\sigma$ constraints on a quiescent radio counterpart to \lptp  at several frequencies. We combined 39 epochs of ASKAP-VAST 12 min observations \citep{2013PASA...30....6M} taken at 887 MHz between 2022 and 2026 for a point-source 3-$\sigma$ limit of 0.20 mJy. We combined 3 epochs of ASKAP-POSSUM observations \citep{2010AAS...21547013G} taken at 1367 MHz in 2022 September for a limit of 70\,$\mu$Jy. We combined 3 epochs of ASKAP-EMU observations \citep{2011PASA...28..215N} taken at 943 MHz for a limit of 80\,$\mu$Jy. Combining the latter three Stokes V images resulted in a limit of 45\,$\mu$Jy for both. We imaged the off-pulse portions of the 2022 May 14 detections for both telescopes;  the lowest limits achieved were with MeerKAT: 70\,$\mu$Jy (Stokes I) and 53\,$\mu$Jy (Stokes V). Finally, combining the former with all archival 1.28 GHz MeerKAT datasets yields a Stokes I limit of 33\,$\mu$Jy.

\startlongtable
\begin{deluxetable}{lllrrr}
\tablecaption{Time-Ordered Summary of Mid-frequency Archival Radio Observations}\label{tab:radio}
\tablehead{\colhead{Obs. Date} & \colhead{Start Time} & \colhead{Telescope} & \colhead{Project ID} & \colhead{Frequency} & \colhead{Time on Source}\\
\colhead{(YY-MM-DD)} & \colhead{(hh:mm)}  & \colhead{} & \colhead{} & \colhead{(MHz)} & \colhead{(min)}
}
\startdata
2019-05-06 & 20:07 & ASKAP & SB8676 beams:01,03&  888 & 15 \\
2019-11-30 & 11:12 & MeerKAT & CB1575109856 & 1279 & 15 \\
2019-12-07 & 09:01 & MeerKAT & CB1575705658 & 1279 & 15 \\
2021-01-30 & 00:15 & ASKAP & SB22178 beams:15& 1368 & 15 \\
2021-07-09 & 23:21 & MeerKAT & CB1625850359 & 1279 & 10 \\
2021-08-10 & 20:46 & MeerKAT & CB1628607085 & 1279 & 10 \\
2021-08-20 & 20:32 & MeerKAT & CB1629471476 & 1279 & 10 \\
2021-09-10 & 18:25 & MeerKAT & CB1631278280 & 1279 & 10 \\
2021-11-06 & 00:37 & ASKAP & SB33284 beams:09,15&  944 & 600 \\
2021-11-08 & 07:58 & MeerKAT & CB1636329974 & 1279 & 10 \\
2021-11-16 & 07:42 & MeerKAT & CB1637020268 & 1279 & 10 \\
2021-11-19 & 07:08 & MeerKAT & CB1637277379 & 1279 & 10 \\
2021-12-16 & 11:47 & MeerKAT & CB1639636576 & 1279 & 10 \\
2021-12-22 & 10:24 & MeerKAT & CB1640150178 & 1279 & 10 \\
2022-01-25 & 00:35 & ASKAP & SB36449 beams:15& 1656 & 16 \\
2022-04-10 & 19:44 & ASKAP & SB39221 beams:01,03&  888 & 15 \\
2022-05-14$^a$ & 11:13 & ASKAP & SB40625 beams:09,15& 1368 & 600 \\
2022-05-14$^a$ & 18:58 & MeerKAT & CB1652551867 & 1279 & 120 \\
2022-06-25 & 16:34 & MeerKAT & CB1656147142 & 1279 & 10 \\
2022-09-07 & 04:22 & ASKAP & SB43773 beams:09,15&  944 & 600 \\
2022-09-13 & 07:28 & ASKAP & SB43990 beams:03& 1368 & 479 \\
2022-09-18 & 02:57 & ASKAP & SB44125 beams:03& 1368 & 480 \\
2022-09-21 & 02:46 & ASKAP & SB44193 beams:02& 1368 & 480 \\
2022-12-07 & 01:44 & ASKAP & SB45989 beams:01,03&  888 & 12 \\
2022-12-08 & 01:31 & ASKAP & SB46045 beams:01,03&  888 & 12 \\
2022-12-22 & 00:57 & ASKAP & SB46895 beams:01,03&  888 & 12 \\
2023-01-05 & 00:03 & ASKAP & SB47019 beams:01,03&  888 & 12 \\
2023-01-20 & 23:01 & ASKAP & SB47245 beams:01,03&  888 & 12 \\
2023-02-03 & 21:40 & ASKAP & SB47618 beams:01,03&  888 & 12 \\
2023-02-04 & 21:39 & ASKAP & SB47683 beams:01,03&  888 & 12 \\
2023-02-18 & 20:46 & ASKAP & SB48252 beams:01,03&  888 & 12 \\
2023-02-22 & 23:05 & ASKAP & SB48455 beams:01,03&  888 & 12 \\
2023-03-06 & 19:35 & ASKAP & SB48844 beams:01,03&  888 & 12 \\
2023-03-22 & 18:35 & ASKAP & SB49007 beams:01,03&  888 & 12 \\
2023-04-06 & 17:38 & ASKAP & SB49139 beams:01,03&  888 & 12 \\
2023-04-21 & 16:35 & ASKAP & SB49586 beams:01,03&  888 & 12 \\
2023-05-06 & 16:00 & ASKAP & SB49865 beams:01,03&  888 & 12 \\
2023-05-21 & 14:51 & ASKAP & SB50142 beams:01,03&  888 & 12 \\
2023-06-02 & 13:50 & ASKAP & SB50303 beams:01,03&  888 & 12 \\
2023-06-05 & 13:40 & ASKAP & SB50383 beams:01,03&  888 & 12 \\
2023-06-12 & 13:53 & ASKAP & SB50442 beams:01,03&  888 & 12 \\
2023-06-21 & 12:57 & ASKAP & SB50641 beams:01,03&  888 & 12 \\
2023-06-22 & 13:59 & ASKAP & SB50690 beams:01,03&  888 & 12 \\
2023-07-01 & 11:17 & ASKAP & SB50966 beams:34,35&  843 & 20 \\
2023-07-04 & 11:45 & ASKAP & SB51095 beams:01,03&  888 & 12 \\
2023-07-21 & 12:42 & ASKAP & SB51525 beams:01,03&  888 & 12 \\
2023-08-16 & 11:38 & ASKAP & SB51975 beams:01,03&  888 & 12 \\
2023-08-31 & 10:03 & ASKAP & SB52405 beams:01,03&  888 & 12 \\
2023-09-16 & 08:10 & ASKAP & SB53205 beams:01,03&  888 & 12 \\
2023-10-04 & 05:37 & ASKAP & SB53487 beams:01,03&  888 & 12 \\
2023-10-19 & 04:36 & ASKAP & SB53904 beams:01,03&  888 & 12 \\
2023-11-02 & 06:04 & ASKAP & SB54719 beams:01,03&  888 & 12 \\
2023-11-19 & 02:37 & ASKAP & SB54910 beams:01,03&  888 & 12 \\
2023-12-31 & 02:14 & ASKAP & SB56373 beams:09,15&  944 & 15 \\
2024-02-09 & 21:50 & ASKAP & SB58915 beams:09,15&  944 & 15 \\
2024-04-16 & 16:46 & ASKAP & SB61156 beams:01,03&  888 & 12 \\
2024-05-01 & 17:53 & ASKAP & SB61955 beams:01,03&  888 & 12 \\
2024-05-20 & 16:46 & ASKAP & SB62485 beams:01,03&  888 & 12 \\
2024-06-07 & 14:05 & ASKAP & SB62635 beams:01,03&  888 & 12 \\
2024-06-23 & 12:48 & ASKAP & SB63227 beams:01,03&  888 & 12 \\
2024-07-13 & 13:07 & ASKAP & SB63644 beams:01,03&  888 & 12 \\
2024-08-14 & 11:08 & ASKAP & SB64759 beams:01,03&  888 & 12 \\
2024-09-05 & 10:18 & ASKAP & SB65646 beams:01,03&  888 & 12 \\
2024-09-28 & 07:09 & ASKAP & SB66121 beams:01,03&  888 & 12 \\
2024-10-21 & 06:52 & ASKAP & SB66903 beams:15& 1368 & 14 \\
2024-10-23 & 04:23 & ASKAP & SB67000 beams:01,03&  888 & 12 \\
2025-06-15 & 10:05 & ASKAP & SB74457 beams:09,15&  944 & 600 \\
2026-02-04 & 23:01 & ASKAP & SB81799 beams:01,03&  888 & 12 \\
2026-02-05 & 00:35 & ASKAP & SB81806 beams:01,03&  888 & 12 \\
2026-02-05 & 20:43 & ASKAP & SB81881 beams:01,03&  888 & 12 \\
2026-02-06 & 20:29 & ASKAP & SB81978 beams:01,03&  888 & 12 \\
2026-02-07 & 21:35 & ASKAP & SB82072 beams:01,03&  888 & 12 \\
2026-02-08 & 22:46 & ASKAP & SB82164 beams:01,03&  888 & 12 \\
\tableline
\enddata
\tablecomments{a. Detections}
\end{deluxetable}

\begin{figure}
    \centering
   \includegraphics[width=\linewidth]{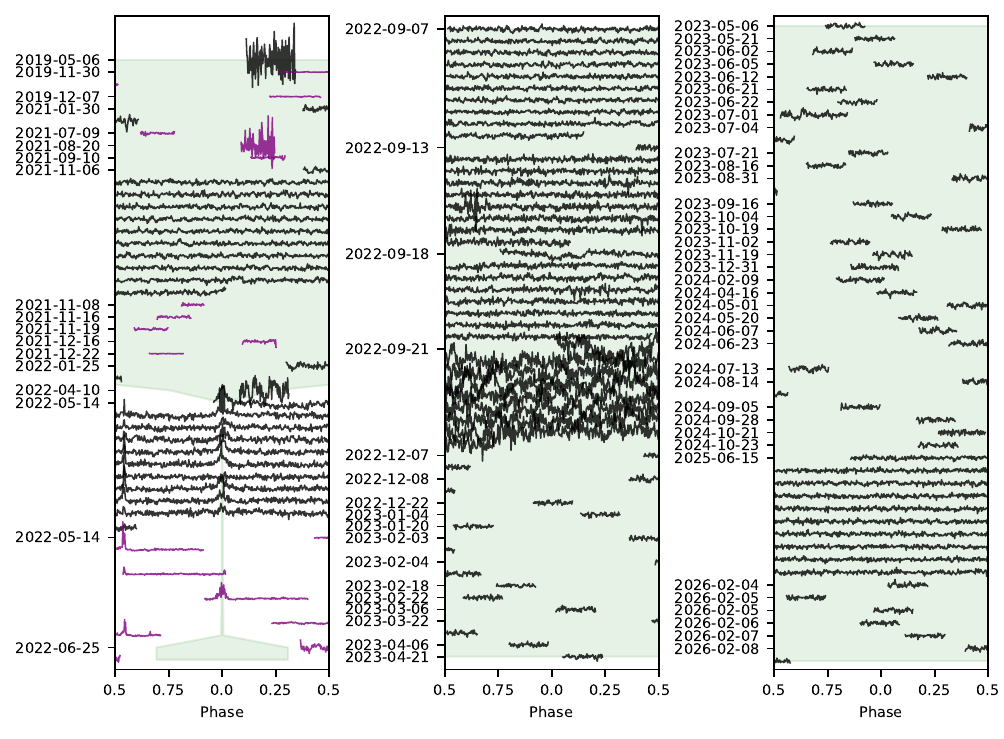}
   \includegraphics[width=\linewidth]{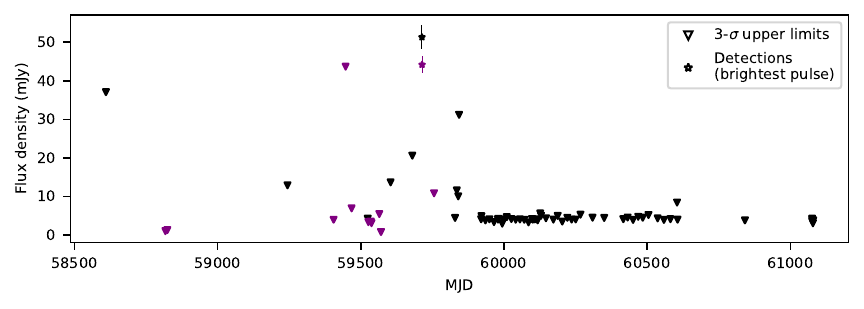}
   \caption{Archival searches. Top panel: all light curves from the archival searches of ASKAP and MeerKAT data, barycentered and folded on the ephemeris in \S\ref{sec:analysis}. Light green shading indicates where the ephemeris is uncertain -- i.e., almost all observations outside of the detections on 2022-05-14. Lower panel: 3-$\sigma$ upper limits from the light curves (frequency-averaged, at their native 10-s (ASKAP) and 8-s (MeerKAT) time resolutions), and the brightest pulses from the two detection observations. ASKAP observations are shown in black and MeerKAT observations are shown in purple. All data are scaled to a common frequency of 1\,GHz via a representative spectral index of $-1.5$.}
    \label{fig:archive_search}
\end{figure}

\section{Multi-Wavelength Data}\label{sec:multi}

\begin{figure}[htb!]
\includegraphics[width=10cm]{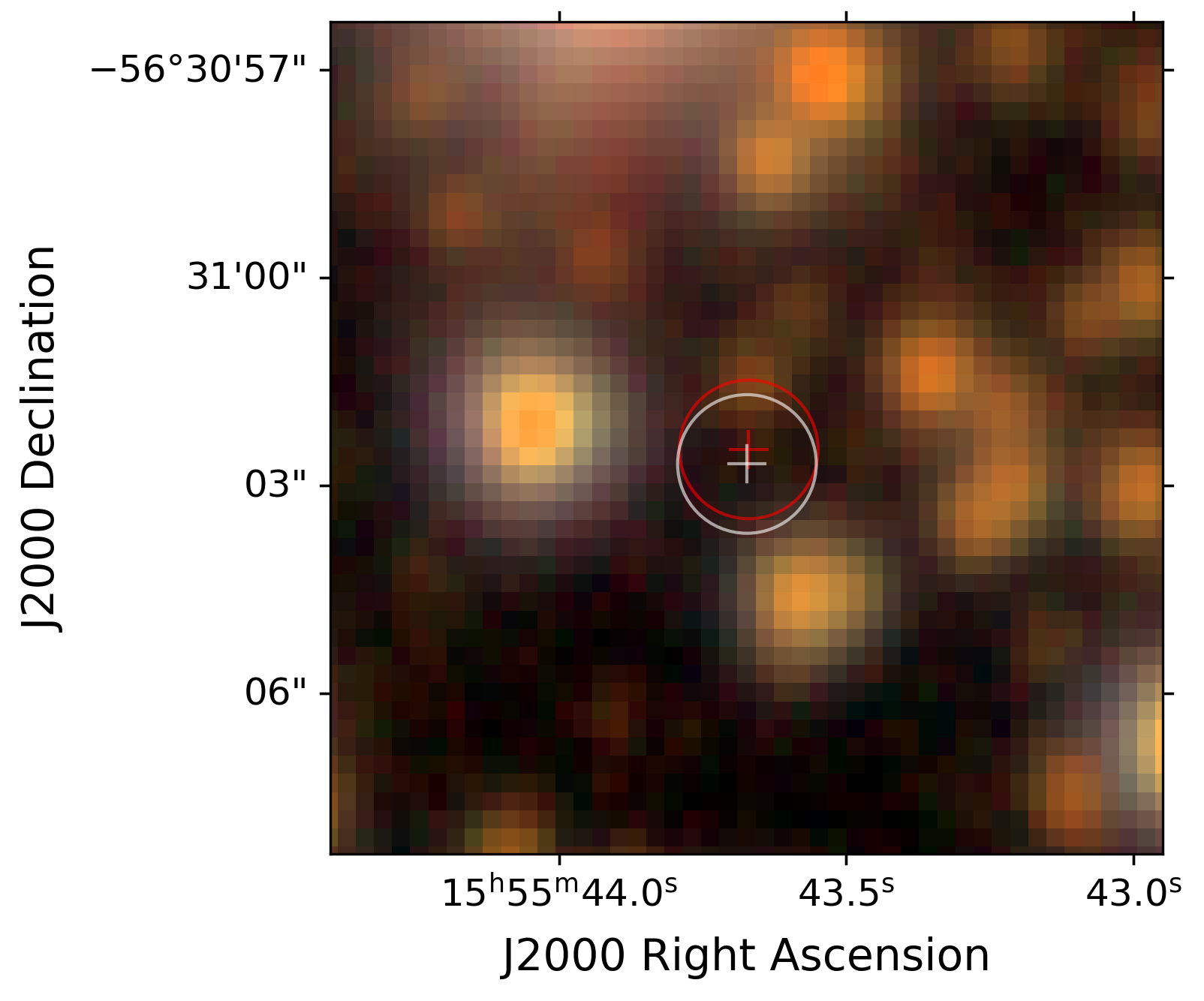}
\centering
\caption{A 3-color postage stamp centered on \lpt made from the DECaPS2 r, i, and z bands. The red and white circles have a 1-arcsec radius centered on the ASKAP and MeerKAT positions, respectively.}\label{fig:DECaPS2}
\end{figure}

We searched for quiescent and variable counterparts of \lpt at optical, infrared and X-ray wavelengths. Optical images at this low Galactic latitude ($l=326.36^\circ$, $b$=-2.27$^\circ$) show a high degree of stellar crowding and prominent dust lanes. Fig. \ref{fig:DECaPS2} is a 3-color composite image around \lpt taken from the Dark Energy Camera Plane Survey 2 \citep[DECaPS2;][]{2023ApJS..264...28S}. The two brightest stars in this DECaPS2 image are also in the latest Gaia catalog \citep{2023A&A...674A...1G}. The closest DECaPS2 source to the north is 1.05$^{\prime\prime}$ and 1.26$^{\prime\prime}$ from \lpt in the MeerKAT and ASKAP images, respectively, and nominally outside of our position uncertainty.

Its  mean magnitudes (Table \ref{table:properties}) and colors are indicative of a nearby M dwarf or a heavily reddened G or K star.  Within a 60$^{\prime\prime}$ of \lpt there are 1462 stars with z$<$20.5, or in other words, we expect 0.45 stars within 1.05$^{\prime\prime}$. The probability of a chance coincidence is high, and we do not claim a physical association.

A search of the ALLWISE catalog from the Wide-Infrared Survey Explorer \citep{2010AJ....140.1868W} also failed to find any emission consistent with \lptp. We likewise searched for any {\it variable} optical or near infrared emission from \lpt using the Vista Variables in the Ví\'a La\'ctea (VVV) survey DR4.2 \citep{2012A&A...537A.107S} and the NEOWISE survey \citep{2011ApJ...743..156M} but there were no cataloged variables within 1-arcmin of this position.

No X-ray matches were found at the position of \lptp. We searched the eRASS1 catalog from the eROSITA telescope \citep{2024A&A...682A..34M} and the XMM-Newton Serendipitous Source Catalog \citep[4XMM-DR-14;][]{2020A&A...641A.136W}. An upper flux limit of 1.5$\times{10}^{-14}$ erg s$^{-1}$ cm$^{-2}$ (0.5 - 7 keV) at this position is given in the Chandra Source Catalog 2.0 \citep{2010ApJS..189...37E}. We used the Web-based PIMMS tool at HEASARC to estimate the X-ray flux corrected for interstellar absorption using two values of the photon index; a pulsar-like hard spectrum ($\Gamma = 1.5$) and a magnetar-like soft spectrum ($\Gamma = 2.0$). The neutral column density (N$_{\rm H}$) in this direction is estimated from 3DN$_{\rm H}$-tool\footnote{\url{http://astro.uni-tuebingen.de/nh3d/nhtool}} to vary from $\leq 2.2\times{10}^{21}$ cm$^{-2}$ below 1 kpc to $\sim$5.6$\times{10}^{21}$ cm$^{-2}$ at 5 kpc \citep{2024arXiv240303127D}. Adopting a nominal distance of d=5.4 kpc (see \S\ref{sec:analysis}) the upper flux limit is 1.9$\times{10}^{-14}$ erg s$^{-1}$ cm$^{-2}$ and 2.3$\times{10}^{-14}$ erg s$^{-1}$ cm$^{-2}$ for $\Gamma$=1.5 and $\Gamma$=2.0, respectively. The degree of correction from N$_{\rm H}$ is neither strongly dependent on the distance nor the value of $\Gamma$. Assuming the same distance, the upper limit on the X-ray luminosity (4$\pi{\rm d}^2 f_x$) would be 7.3$\times{10}^{31}$ erg s$^{-1}$.

\section{Results}\label{sec:results}

\lptp{} was initially discovered in the image plane at L-band as a steep spectrum, circularly polarized source with an anomalous point-spread function (\S\ref{sec:discover}). At higher time resolution the source emits radio pulses with a well-defined period P=62.2 minutes. These pulses were only seen during a 15‑hour window in May 2022. No subsequent detections of pulsations were made in the archival data spanning 2019 to 2026 to a typical 10-sec (ASKAP) and 8-sec (MeerKAT) 3-$\sigma$ level of 4 mJy after scaling to 1 GHz. Upper limits on quiescent radio emission are 33\,$\mu$Jy in Stokes~I and 45\,$\mu$Jy in Stokes~V (\S\ref{sec:archive} and Fig. \ref{fig:archive_search}).

Two distinct alternating pulses were seen from \lpt (\S\ref{sec:analysis} and Fig.~\ref{fig:folded_profiles}), a main pulse (MP) and an interpulse (IP), separated by $\Delta\phi=165.4^\circ$ of phase. While there is significant pulse-to-pulse variation (see Figs. \ref{fig:lightcurves_1030} and \ref{fig:acf}), the amplitude ratio of the {\it mean} pulse profile is IP/MP=1.0$\pm$0.1 (Fig. \ref{fig:folded_profiles} and Table \ref{table:properties}). The main pulse and interpulse each have different polarization properties and pulse profiles (see Fig. \ref{fig:askap_fold} and Table \ref{table:properties}). The MP has a fractional pulse width (or duty cycle) of w50/P=3.9\%. The degree of linear polarization varies in phase across the MP with values from 20\% to 70\%, while the circular polarization varies from 0\% to 90\%. The IP is narrower than the main pulse with a fractional pulse width of 1.1\%. While the linear polarization of the IP is weak ($<$10\%), it is almost 100\% circularly polarized. The circular polarization in the MP and the IP are always of the opposite sign. The polarization position angle of the MP for the mean pulse was measured to be 142$\pm{1}^\circ$ and does not change with pulse phase. Within several bright main pulses we observe regularly spaced sub-pulses superimposed upon several single pulses all with a similar widths and a quasi-periodicity of order 20 to 30 seconds (Fig. \ref{fig:acf}). The wide MeerKAT band enabled us to determine the spectral indices (Fig. \ref{fig:joint_spec}) for two IPs ($\alpha=-2.3, -1.3$) and one MP (-1.3). 

We derived an upper limit on the presence of X-ray emission but were unable to identify an optical counterpart (\S\ref{sec:multi}). There is a faint optical point source just outside the nominal error circle (Fig. \ref{fig:DECaPS2}) but owing to stellar crowding at this low Galactic latitude ($b=-2.3^\circ$) there is a significant non-zero probability of a chance coincidence. We are not able to derive a meaningful DM limit since the expected dispersive delay across the observing bandwidth is small relative to the 62-min period. Likewise, since we lack a long time-baseline, we are unable to measure a period derivative. We determine a rotation measure toward the source of 4.5$\pm$0.7 rad m$^{-2}$. A weak distance constraint of 5.4 kpc is derived by comparing this RM with the average DM-based distances of known pulsars in the vicinity with similar RM values (\S\ref{sec:analysis}).

\section{Discussion and Conclusions}\label{sec:discuss}

\subsection{Classification, Activity Windows and Quiescence}

Collectively, the properties summarized in \S\ref{sec:results} and Table~\ref{table:properties} place \lpt firmly into the phenomenologically-defined class of long period transients. In terms of its period, \lpt is most similar to the LPTs ASKAP J175534.9$-$252749.1 and ASKAP J1935+2148 \citep{2024NatAs...8.1159C,2025MNRAS.542..203M}. In the absence of a robust distance estimate, we can use the radio to X-ray luminosity ratio ($\nu{\rm L}_{\nu,radio}$/L$_X$) to locate \lpt relative to other parent populations \citep{2025Natur.642..583W}. With a mean 1.4 GHz flux density of $\sim$0.5 mJy and $f_x\lesssim$2$\times{10}^{-14}$ erg s$^{-1}$ cm$^{-2}$, \lpt lies along a line where $\nu{\rm L}_{\nu,radio}$/L$_X\leq{3.5\times{10}}^{-4}$. Some caution is warranted as the X-ray and radio data for most LPTs including \lpt are rarely taken concurrently. However, many other LPTs occupy this unfilled phase space with radio to X-ray ratios similar to or lower than \lpt \citep{2025MNRAS.542.1208A,2026ApJ..1003...91A}. LPTs appear radio bright compared to the joint radio and X-ray properties of radio stars, most magnetars and X-ray binaries. For example, the {G{\"u}del}--Benz relationship established for incoherent gyro-synchrotron stars and for luminous (coherent) RS CVn binaries \citep{2022ApJ...926L..30V}, would predict a ratio of $\nu$L$_{\nu,radio}$/L$_X$ at least 2.3 dex lower than our constraint on \lptp. 

The transient nature of the radio emission is widely regarded as a defining characteristic of the LPT class, yet the known population exhibits a surprisingly diverse range of activity cycles \citep{2026JHEAp..5200566R}. While sources such as GLEAM-X\,J0704$-$36 and GPM\,J1839$-$10 have remained active for at least a decade, most LPTs appear to undergo much shorter episodes of radio activity lasting only months. Even during these active intervals the phenomenology varies considerably. Pulse amplitudes may be highly sporadic \citep{2025ApJ...990L..49D}, remain relatively stable \citep{2026arXiv260307857P}, decline steadily with time \citep{2025NatAs...9..393L}, or transition abruptly between distinct emission states \citep{2024NatAs...8.1159C}. Several LPTs also exhibit secondary periodicities that modulate pulse arrival times or their amplitudes \citep{2026NatAs..10..522H,2026ApJ..1003...91A}. These observations suggest that emission is governed intrinsically by changes in LPT magnetospheres or in some cases extrinsically by a binary environment.

Finite activity windows and dormant states are not unique to LPTs. NSs such as magnetars, repeating FRBs, RRATs, intermittent pulsars, and mode-changing pulsars all exhibit long-term changes in radio activity that have been interpreted as transitions between different magnetospheric states \citep{2011BASI...39..333K,2017ARA&A..55..261K,2022ARA&A..60..495P,2025AN....34650024M}. In contrast, the coherent radiation from WD systems such as AR Sco and J1912$-$4410 show regular radio pulses modulated on an orbital period, and accompanied by a persistent low-level component \citep{2016Natur.537..374M,2018A&A...611A..66S,2017A&A...601L...7M}. We are not aware of any WD analog that exhibits the complete disappearance of coherent radio pulses on the timescales inferred for many LPTs.

In this respect, \lpt appears broadly representative of the LPT population. Its radio-loud phase was confined to a short interval lasting at least 15 hr but no more than 10 months (\S\ref{sec:askap_singlepulse}). During this episode, while the pulse-to-pulse amplitude showed variations with an RMS of order 30\%, there is no evidence for the secular fading seen in some LPTs \citep[e.g.,][]{2025NatAs...9..393L}. Outside this interval the source was radio quiet, with no pulses detected over $\sim$5 yr above a typical 3-$\sigma$ threshold of $\sim$4 mJy (\S\ref{sec:askap_singlepulse}). We cannot exclude the possibility that \lpt entered a weaker emission state analogous to GCRT\,J1745$-$3009 \citep{2007ApJ...660L.121H} or the mode switching reported for ASKAP\,J1935+2148 \citep{2024NatAs...8.1159C}, since pulses an order of magnitude fainter than those observed during the active phase would fall below the sensitivity of our archival data (Fig.~\ref{fig:archive_search}). Likewise, our limits on steady, non-pulsed emission (\S\ref{sec:quiet} and Table~\ref{table:properties}) do not yet reach the luminosities measured from AR Sco or J1912$-$4410 unless \lpt lies closer than 0.4 kpc.

\subsection{Magnetospheric Geometry and Emission Processes}

The bright (20-30\,mJy), steep-spectrum ($\alpha\simeq-1.6$) pulses with short duty cycles (1.1--3.9\%) and high degrees of both linear and circular polarization ($>$70\%) from \lpt argue for a coherent emission process such as curvature radiation or the electron cyclotron maser emission, requiring a compact, beamed emission region rather than a distributed, isotropic emitter \citep{2017RvMPP...1....5M}. The detection of a main pulse and an interpulse with opposite circular polarization provides a model-independent geometric constraint. It implies that the emission beam is sampled twice during each cycle. In pulsars, such pulse/interpulse pairs are generally associated with radiation from opposite magnetic poles, requiring a nearly orthogonal geometry between the rotation and magnetic axes (i.e., $\alpha_\circ \simeq90^\circ$). Such objects are called orthogonal rotators \citep{2019MNRAS.490.4565J}. As was first noted by \citet{2025NatAs...9..393L}, if the interpulse interpretation is correct, it implies that the observed period originates from rotation and {\it not} orbital motion. Strong support for this interpretation comes from the sign reversals of the Stokes V signal between the MP and IP (Fig.~\ref{fig:askap_fold}). Among pulsars, \citet{2019MNRAS.490.4565J} have shown that this is a common property of orthogonal rotators, as the line-of-sight component of the magnetic field points toward the observer at one pole and away at the other \citep[e.g, PSR\, B0906$-$49;][]{2008MNRAS.390...87K}. 

\lpt is the second after ASKAP J183950.5$-$075635.0  \citep{2025NatAs...9..393L} to be classified as an orthogonal rotator, or $\sim$15\% of the LPT sample. \lpt and ASKAP J183950.5$-$075635.0, with rotation periods of 62.2 min and 6.45 hrs respectively, demonstrate that orthogonal rotators can be found across nearly the full range of LPT periods. Among the known pulsars this fraction is only 2-3\% \citep{2008MNRAS.387.1755W,2011MNRAS.414.1314M}. Apart from small number statistics, what might explain the greater incidence of orthogonal rotators among LPTs? We argue below that the answer may be that the structure of the emission beam and the location of the emission region differs from canonical pulsar magnetospheres.

Given the orthogonal rotator geometry, we can use the pulse morphology to probe the structure of the emission beam. For most pulsars, the radio emission is thought to arise from the polar cap region relatively low in the magnetosphere, resulting in an opening angle that scales as $\rho=5.8^\circ/\sqrt{\rm{P (s)}}$ \citep{1983ApJ...274..333R}. For \lptp, with $P=3732$ s, the Rankin relation predicts a polar-cap half-angle of only $\rho\approx0.095^\circ$. Instead, the observed duty cycles of 1.1\% (IP) and 3.8\% (MP) imply wider beam half-angles of order $\sim$2--6$^\circ$ (i.e., for a near-orthogonal rotator with $\sin\alpha_\circ\approx1$). This discrepancy of more than an order of magnitude demonstrates that the emission does not originate from a standard polar cap. 

If the emission is from a non-canonical polar-cap beam geometry, what does the MP-IP morphology reveal about the magnetic field topology? In a pure dipole, an orthogonal rotator should produce two anti-podal pulses separated by exactly $\Delta\phi=180^\circ$. The observed separation of $\Delta\phi=165.4^\circ\pm0.2^\circ$ deviates from this ideal by $\Delta\phi_s=14.6^\circ$. For pulsars like the Crab, relativistic aberration and retardation can account for reduced phase separations \citep{1991ApJ...370..643B}, but these effects scale as $\Delta\phi_s\propto v/c = R_{\rm em}/R_{\rm LC}$, where $R_{\rm em}$ is the characteristic radius in the magnetosphere at which the radio emission originates and $R_{\rm LC}=cP/2\pi$ is the light-cylinder radius, while  v is the corotation velocity at $R_{\rm em}$ and c is the speed of light. For \lpt, $R_{\rm LC}\simeq1$ AU, a distance where the magnetic field is too weak to generate or sustain radio emission for either a NS or WD. Thus, if the emission originates much closer to the stellar surface ($R_{\rm em}\ll R_{\rm LC}$), relativistic effects are negligible and cannot explain the $14.6^\circ$ deviation. Instead, the phase offset suggests some sort of asymmetry in the system; a beam asymmetry between the MP and IP emission regions, or geometric deviations from a pure dipole geometry such as an offset dipole or a non-dipolar magnetic field component. The latter may be similar to the twisted field configurations suggested for magnetars \citep{2009ApJ...703.1044B, 2026ApJ...996L..20Z}. Offset dipoles and multipole field topologies are commonly invoked to explain the Zeeman tomography of magnetized WD binaries \citep{2007A&A...463..647B,2018CoSka..48..228K}. Whatever the origin of the asymmetry, it need not be a population-wide property of LPTs since it is not required for the other orthogonal rotator ASKAP J183950.5$-$075635.0 \citep[i.e., $\Delta\phi=177.8^\circ\pm3.0^\circ$;][]{2025NatAs...9..393L}. For future events, stronger constraints can be placed on magnetospheric geometry and the emission regions by looking for temporal changes in $\Delta\phi$
or frequency variations (aka radius-to-frequency mapping).

This hypothesis is partially supported by the polarization position angle (PA) of the MP. The PA was measured to be $142^\circ\pm1^\circ$ and shows no significant variation with pulse phase (Fig.~\ref{fig:askap_fold}).  For a standard pulsar with narrow conal beams, observing both a main pulse and an interpulse with near-equal amplitudes requires a highly improbable, finely-tuned viewing geometry (i.e., one in which the observer's line of sight cuts through the center of both emission regions \citep[e.g.,][]{2011MNRAS.414.1314M, 2019MNRAS.490.4565J}. We discuss a mechanism below that explains the similar IP/MP amplitude without requiring this improbable alignment. Consistent with the atypically wide pulse duty cycles discussed above, we note that a flat PA could arise if the emission region is azimuthally extended, producing a sheet-like or fan beam in which the line of sight samples a nearly constant magnetic longitude \citep{2007MNRAS.380.1678K,2014ApJ...789...73W}. In  \S\ref{sec:population} we discuss the possibility that flat or slowly varying PAs are a population-level property of LPTs.

The high degrees of circular polarization ($>$70\%), the nearly constant polarization PA, and the departure from the standard pulsar polar-cap beam geometry motivate us to consider electron cyclotron maser emission (ECM) as one plausible coherent emission mechanism. ECM is a well-established instability that powers radio emission from cataclysmic variables \citep{2020AdSpR..66.1226B,2026ApJ..1004..123R}, magnetized early-type stars \citep{2018MNRAS.474L..61D}, M dwarfs and brown dwarfs \citep{2019ApJ...871..214V,2022ApJ...932...21K}, and has been invoked for the two known ``WD pulsars'' AR Sco and J1912$-$4410 \citep{2016Natur.537..374M,2023NatAs...7..931P}. ECM differs from coherent curvature radiation in several important respects. Most notably, ECM is emitted approximately perpendicular to the local magnetic field, producing radiation tangential to a thin hollow cone \citep{2019ApJ...877..123D,2024ApJ...974..267D}, whereas coherent curvature radiation is emitted tangentially to the magnetic field lines along the particle trajectory \citep{2017RvMPP...1....5M}. Another key property of ECM is that it grows exponentially but saturates at a maximum brightness temperature when the maser extracts all available energy from the local electron population \citep{1982ApJ...259..844M,2006A&ARv..13..229T}.  Standard coherent curvature radiation does not impose such a hard brightness ceiling, nor does it typically produce such high degrees of circular polarization \citep{2018MNRAS.474.4629J}.

If ECM is indeed the dominant emission mechanism, it may help explain the puzzling properties of the MP and IP of \lptp. The MP ($W_{50}=2.44$ min) is nearly four times wider than the IP ($W_{50}=0.66$ min), yet both reach approximately the same peak flux (IP/MP $\sim$ 1). If both emission regions operate in the saturated regime, the maser intensity is limited by the available free energy in the local electron distribution rather than the size of the emitting region. Two emission components of different pulse widths could therefore exhibit similar peak brightnesses while differing substantially in duration. Our interpretation remains tentative because the mean pulse profile of \lpt is based on only 11 rotations, whereas pulsar mean profiles are typically averaged over hundreds or thousands of pulses. Nevertheless, if future orthogonal-rotator LPTs exhibit similarly different pulse widths but comparable peak brightnesses, it would provide strong support for saturated ECM emission and could provide an observational diagnostic for distinguishing between the two coherent emission mechanisms.

\subsection{Population-Level Comparisons and Future Work}\label{sec:population}

With the geometry, beam structure, and emission mechanism constrained, we look more generally to discriminants of LPT progenitor models, starting with quasi-periodic sub-pulse structure \citep[QPSS;][]{2026arXiv260623970W}. We identified QPSS in three of the brighter pulses from \lpt (\S\ref{sec:single} and Fig \ref{fig:acf}) and measure quasi-periods P$_\mu$=20-30 s.  As the relation predicts a value of $P_\mu=2.7\pm1.1$, \lpt lies significantly above the extrapolation of the $P$--$P_\mu$ relation established by \citet{2024NatAs...8..230K} for NSs. Similar deviations have been reported for other LPTs \citep{2025ApJ...988L..29D}.

Since QPSS appears to be nearly ubiquitous among radio-emitting NSs, its detection in LPT pulses has been argued to provide strong evidence for a NS progenitor \citep{2025NatAs...9..393L,2025ApJ...988L..29D}. However, because the physical origin of QPSS remains poorly understood \citep[][and references therein]{2026arXiv260623970W}, its presence may not be unique to neutron-star magnetospheres \citep{2026JHEAp..5200566R}. Indeed, GPM\,J1839$-$10 illustrates this ambiguity: \citet{2023Natur.619..487H} reported QPSS and used it to place the source on the pulsar $P$--$P_{\nu}$ relation, yet \citet{2025MNRAS.544L..76P} and \citet{2026NatAs..10..522H} have instead argued that the system contains a WD, not a NS. Additionally, QPSS has been claimed from the LPT WD-M star binary GLEAM-X\,J0704$-$37 \citep{2024ApJ...976L..21H}. If more examples of QPSS continue to be found among the WD LPT sub-class, and more are shown to deviate from the $P$--$P_\mu$ relation, then it implies that the magnetospheric plasma microphysics in WDs is capable of producing small-scale emission structure tied to their rotation that is unrelated to QPSS in NSs.

In aggregate, the pulse properties of \lpt favor a magnetized compact object, slowly rotating with a period of 62.2 minutes, in a magnetospheric regime distinct from both standard pulsars and known WD pulsar systems. The detection of a MP and IP with opposite circular polarization implies a near-orthogonal rotator geometry. The large-scale field topology is dipole-like, but a significant multipolar or distorted component (e.g., an offset dipole) is required to account for the 165.4$^\circ$ phase separation and the asymmetric pulse widths. The emission does not originate from a polar cap but from elsewhere in the magnetosphere ($R_{\rm em}\ll R_{\rm LC}$), where it produces broad, saturated ECM beams. This magnetospheric picture explains the origin of the MP and IP, the reversal in the sign of Stokes V between them, their 165.4$^\circ$ phase offset, and their similar mean peak amplitudes. Note that to produce ECM at the first harmonic of the gyro-frequency at the observed 1.4 GHz requires magnetic fields of $\sim$300--500 G \citep{2002ARA&A..40..217G}. Such fields can be achieved within 2--3 WD radii from the surface of a highly magnetized WD ($B\simeq10^9$ G; \citealt{2015SSRv..191..111F}), or alternatively at altitudes of $\sim$10$^4$ km in a NS magnetosphere with $B\sim10^{12}$ G surface field. Thus, ECM is viable for either progenitor.

If this magnetospheric picture applies to LPTs generally, it may explain two intriguing population-level trends: the apparently high incidence of orthogonal rotators and the prevalence of flat polarization position angles. In ordinary pulsars, the Rankin relation ($\rho\propto$P$^{-1/2}$) predicts that long-period objects should have narrow polar-cap beams and therefore rarely be observed as orthogonal rotators \citep{1983ApJ...274..333R,2008MNRAS.387.1755W}. Although the current LPT sample is still small and subject to significant discovery and classification biases, the known population appears to depart from this expectation. LPT duty cycles span $\sim$1.5\%--15\% with no obvious dependence on period \citep{2025MNRAS.542..203M}, and two of the currently identified LPTs have been interpreted as orthogonal rotators \citep{2025NatAs...9..393L}. Whether this reflects a genuine physical difference, small-number statistics, or an observational selection effect remains unclear. If future discoveries confirm these trends, broad, non-canonical polar-cap beam geometry, whether from extended fan-like emission regions or from magnetic interaction zones in binaries would accommodate both the large duty cycles and the high fraction of orthogonal rotators, while removing the expected correlation between beam opening angle and rotation period.

Likewise, flat or slowly varying PAs are equally widespread among LPTs. Sources such as ASKAP\,J075024$-$205945, GLEAM-X J1627$-$5235, and ASKAP\,J1935+218 show little or no PA swing \citep{2022Natur.601..526H, 2024NatAs...8.1159C,2026ApJ..1003...91A}. The two proposed WD systems GPM\,J1839$-$10 and ASKAP\,J1745$-$5051 both have flat PAs but with short-duration phase jumps \citep{2023Natur.619..487H, 2026arXiv260604232R}. The orthogonal rotator ASKAP J183950.5$-$075635.0 exhibited U-shaped polarization swings similar to some pulsars \citep{2025NatAs...9..393L}. A RVM fit was made to ASKAP\,J142431.2$-$612611 but it provided fit parameters atypical of most pulsars and it fails to account for the time-evolution of the circular and linear polarization \citep{2026arXiv260307857P}. 

Two broad classes of explanation have been proposed for these flat PAs. In the first, the PA is determined by the magnetospheric geometry rather than by a dipole field: azimuthally extended (fan-beam) emission regions \citep{2007MNRAS.380.1678K,2014ApJ...789...73W}, twisted or multipolar fields \citep{2007MNRAS.377..107K}, or binary interaction regions \citep{2025ApJ...981...34Q} all produce a nearly constant projected magnetic orientation along the line of sight. In the second, the PA is set by the emission mechanism itself, i.e, saturated electron cyclotron maser emission yields highly beamed, nearly orthogonal polarization that is insensitive to the large-scale field geometry. Both explanations find support in other radio-loud compact objects. Flat PAs are common among FRBs, where the line of sight does not appear to cut through a changing magnetic field configuration \citep{2025ApJ...988..175L}, and among radio-loud magnetars such as XTE\,J1810$-$97,which shows a MP-IP pair with a slowly rising PA \citep{2007ApJ...659L..37C}. Whether LPTs share the magnetar-like non-dipolar fields invoked for XTE\,J1810$-$197 \citep{2007MNRAS.377..107K} or instead represent a distinct electrodynamic regime remains to be determined.

While no single observation conclusively identifies \lpt as either a NS or a WD, its pulse properties serve as a stringent stress-test for models of LPT magnetospheres. The collected radio properties find closer analogs with the NS population (pulsars, RRATs, FRBs, and magnetars), but this may simply reflect the much larger and longer-studied NS population rather than any true physical distinction. The well-known difficulty for a NS model is that for any reasonable value of the B-field, the 62.2-min rotation period places \lpt far beyond the pulsar death line where standard pair-production and coherent radio emission mechanisms are thought to cease \citep{2026JHEAp..5200566R}. As we have shown here, the standard rotation-powered pulsar model also struggles to self-consistently explain all the pulse properties of \lptp. Sustaining beamed radio emission at such long periods is consistent with either a highly twisted, multipolar magnetic field to locally enhance the electric field near the NS surface, or an alternative coherent emission mechanism that does not rely on a polar-cap accelerator \citep{2024MNRAS.533.2133C}.

Conversely, a magnetic WD progenitor accommodates the hour-long rotation period and wide emission beams. However, there has been no detection of coherent emission from {\it isolated} WDs despite sensitive searches \citep[e.g.,][]{2025SCPMA..6829513Z}. Thus the ECM mechanism in WD systems appears to require a binary interaction either through unipolar induction, wind coupling, or accretion to supply the energetic particles and current systems needed for maser amplification \citep{2016ApJ...831L..10G, 2025ApJ...981...34Q,2026ApJ...997..124Y,2026MNRAS.546ag146F}. For \lptp, no companion has been identified, and the 62.2 min spin period is below the $\sim$78 min orbital period minimum for hydrogen-burning, Roche lobe-filling cataclysmic variables \citep{1981ApJ...248L..27P,2009MNRAS.397.2170G}. A detached WD binary with a longer orbital period remains possible, but would require an unseen companion. A similar binary WD model was proposed for GPM\,J1839$-$10 but in this case these was evidence for both a rotation period (P$_r$=22 min) and an orbital period (P$_o$=8.75 hrs) \citep{2025MNRAS.544L..76P,2026NatAs..10..522H}. 

\lpt has shown that regardless of whether the central engine is an ultra-long-period magnetar or a magnetic WD binary, the system operates in a distinct electrodynamic regime compared to known source classes. In this regime, the light cylinder is located at vast distances from the surface, the standard period/opening angle scaling laws break down, and the radio emission geometry is consistent with either local surface multipoles, offset dipoles, or extended binary interaction regions rather than simple dipoles. Several avenues appear promising for future research of \lptp. The first is continued radio monitoring to look for re-activation and to place deeper limits on any quiescent emission. Next, deeper X-ray monitoring (ideally concurrent with radio monitoring) is needed, as well as a proper study of the optical source closest to \lpt to look for any orbital modulation and to better constrain its photometric and spectroscopic properties.

\begin{deluxetable}{l|r}[!ht]
\label{table:properties}
\tabletypesize{\small}
\tablecaption{Observed Properties of \lpt}
\setlength{\tabcolsep}{0.03in}
\tablewidth{0pt}
\tablehead{
\colhead{Property} &
\colhead{Value}
}
\startdata
\hline
\multicolumn{2}{c}{Position} \\
\hline
Right Ascension (J2000, HH:MM:SS) & 15:55:43.67$\pm$0.12 \\
Declination (J2000, DD:MM:SS) & $-$56:31:02.6$\pm$1.0 \\
Galactic longitude ($l$), deg. & 326.36 \\
Galactic latitude ($b$, deg.) & $-$2.27\\
\hline
\multicolumn{2}{c}{Average Pulse Properties} \\
\hline
Period (min) & 62.20486 $\pm$0.0156\\
Reference Epoch (MJD) & 59713.42659906$\pm$0.0000779\\
Main Pulse Amplitude (MP, mJy) & 9.05$\pm$0.28\\
Main pulse width (w50, min) & 2.438$\pm$0.087\\
Main pulse width (w90, min) & 4.444$\pm$0.159\\
Main pulse linear polarization (\%)& 20-70  \\
Main pulse circular polarization (\%) & 0-90  \\
Interpulse Amplitude (IP, mJy) & 9.04$\pm$0.53\\
Interpulse width (w50, min) &  0.663$\pm$0.045\\
Interpulse width (w90, min) &  1.208$\pm$0.082\\
Interpulse linear polarization (\%) & $\leq$10  \\
Interpulse circular polarization (\%) & 100 \\
Main Pulse-Interpulse Separation (deg.) & 165.4$\pm$0.2 \\
Polarization position angle (deg.) & 142$\pm$1\\
\hline
\hline
\multicolumn{2}{c}{Other Parameters} \\
\hline
Rotation Measure (rad m$^{-2}$) & 4.5$\pm$0.7  \\
Active Lifetime & 15 hrs - 10 mo. \\
Persistent flux density ($\mu$Jy) & $<$33 Stokes I; $<$45 Stokes V\\
\hline
\multicolumn{2}{c}{Magnitudes of Nearest Star and X-ray Limits} \\
\hline
g (mag) & 23.00 \\
r (mag) & 21.22 \\
I (mag) & 20.62 \\
z (mag) & 20.12 \\
y (mag) & 19.84 \\
X-ray flux (erg s$^{-1}$ cm$^{-2}$; 0.5 - 7 keV) & $\lesssim$2$\times{10}^{-14}$ \\
\hline
\hline
\enddata
\tablecomments{}
\end{deluxetable}

\section*{Acknowledgements}
The authors would like to thank Alec Thompson and Emil Lenc for extensive guidance on ASKAP polarimetry, and Josh Pritchard for kindly sharing his light curve generation code, used as the basis for \Fig~\ref{fig:archive_search}. Thanks to Csan{\'a}d Horv{\'a}th, Andrew Zic, and Ziteng Wang for discussions on the radio properties of LPTs.

The MeerKAT telescope is operated by the South African Radio Astronomy Observatory, which is a facility of the National Research Foundation, an agency of the Department of Science and Innovation.

This scientific work uses data obtained from Inyarrimanha Ilgari Bundara, the CSIRO Murchison Radio-astronomy Observatory. We acknowledge the Wajarri Yamaji People as the Traditional Owners and native title holders of the Observatory site. CSIRO’s ASKAP radio telescope is part of the Australia Telescope National Facility\footnote{\url{https://ror.org/05qajvd42}}. Operation of ASKAP is funded by the Australian Government with support from the National Collaborative Research Infrastructure Strategy. ASKAP uses the resources of the Pawsey Supercomputing Research Centre. Establishment of ASKAP, Inyarrimanha Ilgari Bundara, the CSIRO Murchison Radio-astronomy Observatory and the Pawsey Supercomputing Research Centre are initiatives of the Australian Government, with support from the Government of Western Australia and the Science and Industry Endowment Fund.

Parts of the results in this work make use of the colormaps in the CMasher package \citep{cmasher}. Basic research at NRL is funded by 6.1 Base programs. 

This research has made use of data and/or software provided by the High Energy Astrophysics Science Archive Research Center (HEASARC), which is a service of the Astrophysics Science Division at NASA/GSFC.

\vspace{5mm}
\facilities{MeerKAT, ASKAP}

\software{PyBDSF \citep{2015ascl.soft02007M}, astropy \citep{2013A&A...558A..33A,2018AJ....156..123A},  
Obit \citep{2008PASP..120..439C}, 
CASA \citep{2022PASP..134k4501C},
and AIPS \citep{2003ASSL..285..109G}
}

\section*{Data Availability}
The MeerKAT and ASKAP dynamic spectra are provided at Zenodo: \dataset[doi:10.5281/zenodo.22822831]{https://doi.org/10.5281/zenodo.22822831}. The code used for analysis and to produce the figures used in this paper can be found on GitHub at \url{https://github.com/nhurleywalker/LPTJ1555-56}.

\bibliography{polarcandi}{}

@ARTICLE{2006ApJ...639..348H,
       author = {{Hyman}, Scott D. and {Lazio}, T. Joseph W. and {Roy}, Subhashis and {Ray}, Paul S. and {Kassim}, Namir E. and {Neureuther}, Jennifer L.},
        title = "{A New Radio Detection of the Transient Bursting Source GCRT J1745-3009}",
      journal = {\apj},
         year = 2006,
        month = mar,
       volume = {639},
       number = {1},
        pages = {348-353},
          doi = {10.1086/499294},
archivePrefix = {arXiv},
       eprint = {astro-ph/0508264},
 primaryClass = {astro-ph},
       adsurl = {https://ui.adsabs.harvard.edu/abs/2006ApJ...639..348H}
}

@ARTICLE{1981ApJ...248L..27P,
       author = {{Paczynski}, B. and {Sienkiewicz}, R.},
        title = "{Gravitational radiation and the evolution of cataclysmic binaries}",
      journal = {\apjl},
         year = 1981,
        month = aug,
       volume = {248},
        pages = {L27-L30},
          doi = {10.1086/183616},
       adsurl = {https://ui.adsabs.harvard.edu/abs/1981ApJ...248L..27P}
}

@ARTICLE{2009MNRAS.397.2170G,
       author = {{G{\"a}nsicke}, B.~T. and {Dillon}, M. and {Southworth}, J. and {Thorstensen}, J.~R. and {Rodr{\'\i}guez-Gil}, P. and {Aungwerojwit}, A. and {Marsh}, T.~R. and {Szkody}, P. and {Barros}, S.~C.~C. and {Casares}, J. and {de Martino}, D. and {Groot}, P.~J. and {Hakala}, P. and {Kolb}, U. and {Littlefair}, S.~P. and {Mart{\'\i}nez-Pais}, I.~G. and {Nelemans}, G. and {Schreiber}, M.~R.},
        title = "{SDSS unveils a population of intrinsically faint cataclysmic variables at the minimum orbital period}",
      journal = {\mnras},
         year = 2009,
        month = aug,
       volume = {397},
       number = {4},
        pages = {2170-2188},
          doi = {10.1111/j.1365-2966.2009.15126.x},
archivePrefix = {arXiv},
       eprint = {0905.3476},
 primaryClass = {astro-ph.SR},
       adsurl = {https://ui.adsabs.harvard.edu/abs/2009MNRAS.397.2170G}
}

@ARTICLE{2023ApJS..264...28S,
       author = {{Saydjari}, Andrew K. and {Schlafly}, Edward F. and {Lang}, Dustin and {Meisner}, Aaron M. and {Green}, Gregory M. and {Zucker}, Catherine and {Zelko}, Ioana and {Speagle}, Joshua S. and {Daylan}, Tansu and {Lee}, Albert and {Valdes}, Francisco and {Schlegel}, David and {Finkbeiner}, Douglas P.},
        title = "{The Dark Energy Camera Plane Survey 2 (DECaPS2): More Sky, Less Bias, and Better Uncertainties}",
      journal = {\apjs},
         year = 2023,
        month = feb,
       volume = {264},
       number = {2},
          eid = {28},
        pages = {28},
          doi = {10.3847/1538-4365/aca594},
archivePrefix = {arXiv},
       eprint = {2206.11909},
 primaryClass = {astro-ph.GA},
       adsurl = {https://ui.adsabs.harvard.edu/abs/2023ApJS..264...28S}
}

@ARTICLE{2020A&A...641A.136W,
       author = {{Webb}, N.~A. and {Coriat}, M. and {Traulsen}, I. and {Ballet}, J. and {Motch}, C. and {Carrera}, F.~J. and {Koliopanos}, F. and {Authier}, J. and {de la Calle}, I. and {Ceballos}, M.~T. and {Colomo}, E. and {Chuard}, D. and {Freyberg}, M. and {Garcia}, T. and {Kolehmainen}, M. and {Lamer}, G. and {Lin}, D. and {Maggi}, P. and {Michel}, L. and {Page}, C.~G. and {Page}, M.~J. and {Perea-Calderon}, J.~V. and {Pineau}, F.-X. and {Rodriguez}, P. and {Rosen}, S.~R. and {Santos Lleo}, M. and {Saxton}, R.~D. and {Schwope}, A. and {Tom{\'a}s}, L. and {Watson}, M.~G. and {Zakardjian}, A.},
        title = "{The XMM-Newton serendipitous survey. IX. The fourth XMM-Newton serendipitous source catalogue}",
      journal = {\aap},
         year = 2020,
        month = sep,
       volume = {641},
          eid = {A136},
        pages = {A136},
          doi = {10.1051/0004-6361/201937353},
archivePrefix = {arXiv},
       eprint = {2007.02899},
 primaryClass = {astro-ph.HE},
       adsurl = {https://ui.adsabs.harvard.edu/abs/2020A&A...641A.136W}
}

@ARTICLE{2022ApJ...935..167A,
       author = {{Astropy Collaboration} and {Price-Whelan}, Adrian M. and {Lim}, Pey Lian and {Earl}, Nicholas and {Starkman}, Nathaniel and {Bradley}, Larry and {Shupe}, David L. and {Patil}, Aarya A. and {Corrales}, Lia and {Brasseur}, C.~E. and {N{\"o}the}, Maximilian and {Donath}, Axel and {Tollerud}, Erik and {Morris}, Brett M. and {Ginsburg}, Adam and {Vaher}, Eero and {Weaver}, Benjamin A. and {Tocknell}, James and {Jamieson}, William and {van Kerkwijk}, Marten H. and {Robitaille}, Thomas P. and {Merry}, Bruce and {Bachetti}, Matteo and {G{\"u}nther}, H. Moritz and {Aldcroft}, Thomas L. and {Alvarado-Montes}, Jaime A. and {Archibald}, Anne M. and {B{\'o}di}, Attila and {Bapat}, Shreyas and {Barentsen}, Geert and {Baz{\'a}n}, Juanjo and {Biswas}, Manish and {Boquien}, M{\'e}d{\'e}ric and {Burke}, D.~J. and {Cara}, Daria and {Cara}, Mihai and {Conroy}, Kyle E. and {Conseil}, Simon and {Craig}, Matthew W. and {Cross}, Robert M. and {Cruz}, Kelle L. and {D'Eugenio}, Francesco and {Dencheva}, Nadia and {Devillepoix}, Hadrien A.~R. and {Dietrich}, J{\"o}rg P. and {Eigenbrot}, Arthur Davis and {Erben}, Thomas and {Ferreira}, Leonardo and {Foreman-Mackey}, Daniel and {Fox}, Ryan and {Freij}, Nabil and {Garg}, Suyog and {Geda}, Robel and {Glattly}, Lauren and {Gondhalekar}, Yash and {Gordon}, Karl D. and {Grant}, David and {Greenfield}, Perry and {Groener}, Austen M. and {Guest}, Steve and {Gurovich}, Sebastian and {Handberg}, Rasmus and {Hart}, Akeem and {Hatfield-Dodds}, Zac and {Homeier}, Derek and {Hosseinzadeh}, Griffin and {Jenness}, Tim and {Jones}, Craig K. and {Joseph}, Prajwel and {Kalmbach}, J. Bryce and {Karamehmetoglu}, Emir and {Ka{\l}uszy{\'n}ski}, Miko{\l}aj and {Kelley}, Michael S.~P. and {Kern}, Nicholas and {Kerzendorf}, Wolfgang E. and {Koch}, Eric W. and {Kulumani}, Shankar and {Lee}, Antony and {Ly}, Chun and {Ma}, Zhiyuan and {MacBride}, Conor and {Maljaars}, Jakob M. and {Muna}, Demitri and {Murphy}, N.~A. and {Norman}, Henrik and {O'Steen}, Richard and {Oman}, Kyle A. and {Pacifici}, Camilla and {Pascual}, Sergio and {Pascual-Granado}, J. and {Patil}, Rohit R. and {Perren}, Gabriel I. and {Pickering}, Timothy E. and {Rastogi}, Tanuj and {Roulston}, Benjamin R. and {Ryan}, Daniel F. and {Rykoff}, Eli S. and {Sabater}, Jose and {Sakurikar}, Parikshit and {Salgado}, Jes{\'u}s and {Sanghi}, Aniket and {Saunders}, Nicholas and {Savchenko}, Volodymyr and {Schwardt}, Ludwig and {Seifert-Eckert}, Michael and {Shih}, Albert Y. and {Jain}, Anany Shrey and {Shukla}, Gyanendra and {Sick}, Jonathan and {Simpson}, Chris and {Singanamalla}, Sudheesh and {Singer}, Leo P. and {Singhal}, Jaladh and {Sinha}, Manodeep and {Sip{\H{o}}cz}, Brigitta M. and {Spitler}, Lee R. and {Stansby}, David and {Streicher}, Ole and {{\v{S}}umak}, Jani and {Swinbank}, John D. and {Taranu}, Dan S. and {Tewary}, Nikita and {Tremblay}, Grant R. and {de Val-Borro}, Miguel and {Van Kooten}, Samuel J. and {Vasovi{\'c}}, Zlatan and {Verma}, Shresth and {de Miranda Cardoso}, Jos{\'e} Vin{\'\i}cius and {Williams}, Peter K.~G. and {Wilson}, Tom J. and {Winkel}, Benjamin and {Wood-Vasey}, W.~M. and {Xue}, Rui and {Yoachim}, Peter and {Zhang}, Chen and {Zonca}, Andrea and {Astropy Project Contributors}},
        title = "{The Astropy Project: Sustaining and Growing a Community-oriented Open-source Project and the Latest Major Release (v5.0) of the Core Package}",
      journal = {\apj},
         year = 2022,
        month = aug,
       volume = {935},
       number = {2},
          eid = {167},
        pages = {167},
          doi = {10.3847/1538-4357/ac7c74},
archivePrefix = {arXiv},
       eprint = {2206.14220},
 primaryClass = {astro-ph.IM},
       adsurl = {https://ui.adsabs.harvard.edu/abs/2022ApJ...935..167A}
}

@ARTICLE{2021PASA...38...46N,
       author = {{Norris}, Ray P. and {Marvil}, Joshua and {Collier}, J.~D. and {Kapi{\'n}ska}, Anna D. and {O'Brien}, Andrew N. and {Rudnick}, L. and {Andernach}, Heinz and {Asorey}, Jacobo and {Brown}, Michael J.~I. and {Br{\"u}ggen}, Marcus and {Crawford}, Evan and {English}, Jayanne and {Rahman}, Syed Faisal ur and {Filipovi{\'c}}, Miroslav D. and {Gordon}, Yjan and {G{\"u}rkan}, G{\"u}lay and {Hale}, Catherine and {Hopkins}, Andrew M. and {Huynh}, Minh T. and {HyeongHan}, Kim and {James Jee}, M. and {Koribalski}, B{\"a}rbel S. and {Lenc}, Emil and {Luken}, Kieran and {Parkinson}, David and {Prandoni}, Isabella and {Raja}, Wasim and {Reiprich}, Thomas H. and {Riseley}, Christopher J. and {Shabala}, Stanislav S. and {Sheil}, Jaimie R. and {Vernstrom}, Tessa and {Whiting}, Matthew T. and {Allison}, James R. and {Anderson}, C.~S. and {Ball}, Lewis and {Bell}, Martin and {Bunton}, John and {Galvin}, T.~J. and {Gupta}, Neeraj and {Hotan}, Aidan and {Jacka}, Colin and {Macgregor}, Peter J. and {Mahony}, Elizabeth K. and {Maio}, Umberto and {Moss}, Vanessa and {Pandey-Pommier}, M. and {Voronkov}, Maxim A.},
        title = "{The Evolutionary Map of the Universe pilot survey}",
      journal = {\pasa},
         year = 2021,
        month = sep,
       volume = {38},
          eid = {e046},
        pages = {e046},
          doi = {10.1017/pasa.2021.42},
archivePrefix = {arXiv},
       eprint = {2108.00569},
 primaryClass = {astro-ph.CO},
       adsurl = {https://ui.adsabs.harvard.edu/abs/2021PASA...38...46N}
}

@INPROCEEDINGS{1986syim.conf..123C,
       author = {{Cotton}, W.~D.},
        title = "{Special problems in imaging.}",
    booktitle = {Synthesis Imaging},
         year = 1986,
       editor = {{Perley}, Richard A. and {Schwab}, F.~R. and {Bridle}, Alan H.},
        month = jan,
        pages = {123-136},
       adsurl = {https://ui.adsabs.harvard.edu/abs/1986syim.conf..123C}
}

@ARTICLE{2011ApJ...743..156M,
       author = {{Mainzer}, A. and {Grav}, T. and {Bauer}, J. and {Masiero}, J. and {McMillan}, R.~S. and {Cutri}, R.~M. and {Walker}, R. and {Wright}, E. and {Eisenhardt}, P. and {Tholen}, D.~J. and {Spahr}, T. and {Jedicke}, R. and {Denneau}, L. and {DeBaun}, E. and {Elsbury}, D. and {Gautier}, T. and {Gomillion}, S. and {Hand}, E. and {Mo}, W. and {Watkins}, J. and {Wilkins}, A. and {Bryngelson}, G.~L. and {Del Pino Molina}, A. and {Desai}, S. and {G{\'o}mez Camus}, M. and {Hidalgo}, S.~L. and {Konstantopoulos}, I. and {Larsen}, J.~A. and {Maleszewski}, C. and {Malkan}, M.~A. and {Mauduit}, J.-C. and {Mullan}, B.~L. and {Olszewski}, E.~W. and {Pforr}, J. and {Saro}, A. and {Scotti}, J.~V. and {Wasserman}, L.~H.},
        title = "{NEOWISE Observations of Near-Earth Objects: Preliminary Results}",
      journal = {\apj},
         year = 2011,
        month = dec,
       volume = {743},
       number = {2},
          eid = {156},
        pages = {156},
          doi = {10.1088/0004-637X/743/2/156},
archivePrefix = {arXiv},
       eprint = {1109.6400},
 primaryClass = {astro-ph.EP},
       adsurl = {https://ui.adsabs.harvard.edu/abs/2011ApJ...743..156M}
}

@ARTICLE{2026arXiv260604232R,
       author = {{Rose}, Kovi and {Pritchard}, Joshua and {Murphy}, Tara and {Driessen}, L.~N. and {Kaplan}, D.~L. and {Caleb}, M. and {Wang}, Ziteng and {Zic}, A. and {Andreoni}, I. and {Carney}, J. and {Barlow}, B.~N. and {Dobie}, D. and {Gu}, M. and {Heald}, G. and {Huber}, D. and {Lenc}, E. and {Leung}, J.~K. and {Lu}, W. and {Momose}, R. and {Pedersen}, M.~G. and {Qu}, Y. and {Rea}, N. and {de Ruiter}, I. and {Shaji}, K. and {Sivakoff}, G.~R. and {Thomson}, A.~J.~M. and {Wang}, Y.~L. and {Yang}, G.~J. and {Zahedy}, F.},
        title = "{Periodic Radio and X-ray Emission from an Accreting White Dwarf Binary}",
      journal = {arXiv e-prints},
         year = 2026,
        month = jun,
          eid = {arXiv:2606.04232},
        pages = {arXiv:2606.04232},
archivePrefix = {arXiv},
       eprint = {2606.04232},
 primaryClass = {astro-ph.HE},
       adsurl = {https://ui.adsabs.harvard.edu/abs/2026arXiv260604232R}
}

@ARTICLE{2012A&A...537A.107S,
       author = {{Saito}, R.~K. and {Hempel}, M. and {Minniti}, D. and {Lucas}, P.~W. and {Rejkuba}, M. and {Toledo}, I. and {Gonzalez}, O.~A. and {Alonso-Garc{\'\i}a}, J. and {Irwin}, M.~J. and {Gonzalez-Solares}, E. and {Hodgkin}, S.~T. and {Lewis}, J.~R. and {Cross}, N. and {Ivanov}, V.~D. and {Kerins}, E. and {Emerson}, J.~P. and {Soto}, M. and {Am{\^o}res}, E.~B. and {Gurovich}, S. and {D{\'e}k{\'a}ny}, I. and {Angeloni}, R. and {Beamin}, J.~C. and {Catelan}, M. and {Padilla}, N. and {Zoccali}, M. and {Pietrukowicz}, P. and {Moni Bidin}, C. and {Mauro}, F. and {Geisler}, D. and {Folkes}, S.~L. and {Sale}, S.~E. and {Borissova}, J. and {Kurtev}, R. and {Ahumada}, A.~V. and {Alonso}, M.~V. and {Adamson}, A. and {Arias}, J.~I. and {Bandyopadhyay}, R.~M. and {Barb{\'a}}, R.~H. and {Barbuy}, B. and {Baume}, G.~L. and {Bedin}, L.~R. and {Bellini}, A. and {Benjamin}, R. and {Bica}, E. and {Bonatto}, C. and {Bronfman}, L. and {Carraro}, G. and {Chen{\`e}}, A.~N. and {Clari{\'a}}, J.~J. and {Clarke}, J.~R.~A. and {Contreras}, C. and {Corvill{\'o}n}, A. and {de Grijs}, R. and {Dias}, B. and {Drew}, J.~E. and {Fari{\~n}a}, C. and {Feinstein}, C. and {Fern{\'a}ndez-Laj{\'u}s}, E. and {Gamen}, R.~C. and {Gieren}, W. and {Goldman}, B. and {Gonz{\'a}lez-Fern{\'a}ndez}, C. and {Grand}, R.~J.~J. and {Gunthardt}, G. and {Hambly}, N.~C. and {Hanson}, M.~M. and {He{\l}miniak}, K.~G. and {Hoare}, M.~G. and {Huckvale}, L. and {Jord{\'a}n}, A. and {Kinemuchi}, K. and {Longmore}, A. and {L{\'o}pez-Corredoira}, M. and {Maccarone}, T. and {Majaess}, D. and {Mart{\'\i}n}, E.~L. and {Masetti}, N. and {Mennickent}, R.~E. and {Mirabel}, I.~F. and {Monaco}, L. and {Morelli}, L. and {Motta}, V. and {Palma}, T. and {Parisi}, M.~C. and {Parker}, Q. and {Pe{\~n}aloza}, F. and {Pietrzy{\'n}ski}, G. and {Pignata}, G. and {Popescu}, B. and {Read}, M.~A. and {Rojas}, A. and {Roman-Lopes}, A. and {Ruiz}, M.~T. and {Saviane}, I. and {Schreiber}, M.~R. and {Schr{\"o}der}, A.~C. and {Sharma}, S. and {Smith}, M.~D. and {Sodr{\'e}}, L. and {Stead}, J. and {Stephens}, A.~W. and {Tamura}, M. and {Tappert}, C. and {Thompson}, M.~A. and {Valenti}, E. and {Vanzi}, L. and {Walton}, N.~A. and {Weidmann}, W. and {Zijlstra}, A.},
        title = "{VVV DR1: The first data release of the Milky Way bulge and southern plane from the near-infrared ESO public survey VISTA variables in the V{\'\i}a L{\'a}ctea}",
      journal = {\aap},
         year = 2012,
        month = jan,
       volume = {537},
          eid = {A107},
        pages = {A107},
          doi = {10.1051/0004-6361/201118407},
archivePrefix = {arXiv},
       eprint = {1111.5511},
 primaryClass = {astro-ph.GA},
       adsurl = {https://ui.adsabs.harvard.edu/abs/2012A&A...537A.107S}
}

@ARTICLE{2011PASA...28..215N,
       author = {{Norris}, Ray P. and {Hopkins}, A.~M. and {Afonso}, J. and {Brown}, S. and {Condon}, J.~J. and {Dunne}, L. and {Feain}, I. and {Hollow}, R. and {Jarvis}, M. and {Johnston-Hollitt}, M. and {Lenc}, E. and {Middelberg}, E. and {Padovani}, P. and {Prandoni}, I. and {Rudnick}, L. and {Seymour}, N. and {Umana}, G. and {Andernach}, H. and {Alexander}, D.~M. and {Appleton}, P.~N. and {Bacon}, D. and {Banfield}, J. and {Becker}, W. and {Brown}, M.~J.~I. and {Ciliegi}, P. and {Jackson}, C. and {Eales}, S. and {Edge}, A.~C. and {Gaensler}, B.~M. and {Giovannini}, G. and {Hales}, C.~A. and {Hancock}, P. and {Huynh}, M.~T. and {Ibar}, E. and {Ivison}, R.~J. and {Kennicutt}, R. and {Kimball}, Amy E. and {Koekemoer}, A.~M. and {Koribalski}, B.~S. and {L{\'o}pez-S{\'a}nchez}, {\'A}. R. and {Mao}, M.~Y. and {Murphy}, T. and {Messias}, H. and {Pimbblet}, K.~A. and {Raccanelli}, A. and {Randall}, K.~E. and {Reiprich}, T.~H. and {Roseboom}, I.~G. and {R{\"o}ttgering}, H. and {Saikia}, D.~J. and {Sharp}, R.~G. and {Slee}, O.~B. and {Smail}, Ian and {Thompson}, M.~A. and {Urquhart}, J.~S. and {Wall}, J.~V. and {Zhao}, G.-B.},
        title = "{EMU: Evolutionary Map of the Universe}",
      journal = {\pasa},
         year = 2011,
        month = aug,
       volume = {28},
       number = {3},
        pages = {215-248},
          doi = {10.1071/AS11021},
archivePrefix = {arXiv},
       eprint = {1106.3219},
 primaryClass = {astro-ph.CO},
       adsurl = {https://ui.adsabs.harvard.edu/abs/2011PASA...28..215N}
}

@ARTICLE{2017ApJ...851..128T,
       author = {{Temim}, Tea and {Slane}, Patrick and {Plucinsky}, Paul P. and {Gelfand}, Joseph and {Castro}, Daniel and {Kolb}, Christopher},
        title = "{Proper Motion of the High-velocity Pulsar in SNR MSH 15-56}",
      journal = {\apj},
         year = 2017,
        month = dec,
       volume = {851},
       number = {2},
          eid = {128},
        pages = {128},
          doi = {10.3847/1538-4357/aa9d41},
archivePrefix = {arXiv},
       eprint = {1711.09067},
 primaryClass = {astro-ph.HE},
       adsurl = {https://ui.adsabs.harvard.edu/abs/2017ApJ...851..128T}
}

@ARTICLE{2026JHEAp..5200566R,
       author = {{Rea}, Nanda and {Hurley-Walker}, Natasha and {Caleb}, Manisha},
        title = "{Long period transients (LPTs): A comprehensive review}",
      journal = {Journal of High Energy Astrophysics},
         year = 2026,
        month = apr,
       volume = {52},
          eid = {100566},
        pages = {100566},
          doi = {10.1016/j.jheap.2026.100566},
archivePrefix = {arXiv},
       eprint = {2601.10393},
 primaryClass = {astro-ph.HE},
       adsurl = {https://ui.adsabs.harvard.edu/abs/2026JHEAp..5200566R}
}

@ARTICLE{2024ApJS..270...21C,
       author = {{Cotton}, W.~D. and {Kothes}, R. and {Camilo}, F. and {Chandra}, P. and {Buchner}, S. and {Nyamai}, M.},
        title = "{MeerKAT 1.3 GHz Observations of Supernova Remnants}",
      journal = {\apjs},
         year = 2024,
        month = feb,
       volume = {270},
       number = {2},
          eid = {21},
        pages = {21},
          doi = {10.3847/1538-4365/ad0ecb},
archivePrefix = {arXiv},
       eprint = {2311.12140},
 primaryClass = {astro-ph.HE},
       adsurl = {https://ui.adsabs.harvard.edu/abs/2024ApJS..270...21C}
}

@ARTICLE{2026arXiv260516917T,
       author = {{Thomson}, Alec J.~M. and {Galvin}, Timothy J. and {Duchesne}, Stefan W. and {Lenc}, Emil and {Heald}, George and {Hlinka}, Ondrej and {Malik}, Sunil and {Anderson}, Craig S. and {Osinga}, Erik and {Baidoo}, Lerato and {McClure-Griffiths}, N.~M. and {Hutschenreuter}, Sebastian and {O'Sullivan}, Shane P. and {Akahori}, Takuya and {Gaensler}, B.~M. and {Leahy}, J.~P. and {Ma}, Y.~K. and {Moss}, Vanessa A. and {Rudnick}, L. and {Van Eck}, C.~L. and {West}, J.~L.},
        title = "{The Rapid ASKAP Continuum Survey VII: Spectra and Polarisation In Cutouts of Extragalactic Sources (SPICE-RACS) Second Data Release -- Unveiling the Magnetised Sky}",
      journal = {arXiv e-prints},
         year = 2026,
        month = may,
          eid = {arXiv:2605.16917},
        pages = {arXiv:2605.16917},
          doi = {10.48550/arXiv.2605.16917},
archivePrefix = {arXiv},
       eprint = {2605.16917},
 primaryClass = {astro-ph.GA},
       adsurl = {https://ui.adsabs.harvard.edu/abs/2026arXiv260516917T}
}

@INPROCEEDINGS{2003ASSL..285..109G,
       author = {{Greisen}, E.~W.},
        title = "{AIPS, the VLA, and the VLBA}",
    booktitle = {Information Handling in Astronomy - Historical Vistas},
         year = 2003,
       editor = {{Heck}, Andr{\'e}},
       series = {Astrophysics and Space Science Library},
       volume = {285},
        month = mar,
        pages = {109},
          doi = {10.1007/0-306-48080-8_7},
       adsurl = {https://ui.adsabs.harvard.edu/abs/2003ASSL..285..109G}
}

@ARTICLE{2022PASP..134k4501C,
       author = {{CASA Team} and {Bean}, Ben and {Bhatnagar}, Sanjay and {Castro}, Sandra and {Donovan Meyer}, Jennifer and {Emonts}, Bjorn and {Garcia}, Enrique and {Garwood}, Robert and {Golap}, Kumar and {Gonzalez Villalba}, Justo and {Harris}, Pamela and {Hayashi}, Yohei and {Hoskins}, Josh and {Hsieh}, Mingyu and {Jagannathan}, Preshanth and {Kawasaki}, Wataru and {Keimpema}, Aard and {Kettenis}, Mark and {Lopez}, Jorge and {Marvil}, Joshua and {Masters}, Joseph and {McNichols}, Andrew and {Mehringer}, David and {Miel}, Renaud and {Moellenbrock}, George and {Montesino}, Federico and {Nakazato}, Takeshi and {Ott}, Juergen and {Petry}, Dirk and {Pokorny}, Martin and {Raba}, Ryan and {Rau}, Urvashi and {Schiebel}, Darrell and {Schweighart}, Neal and {Sekhar}, Srikrishna and {Shimada}, Kazuhiko and {Small}, Des and {Steeb}, Jan-Willem and {Sugimoto}, Kanako and {Suoranta}, Ville and {Tsutsumi}, Takahiro and {van Bemmel}, Ilse M. and {Verkouter}, Marjolein and {Wells}, Akeem and {Xiong}, Wei and {Szomoru}, Arpad and {Griffith}, Morgan and {Glendenning}, Brian and {Kern}, Jeff},
        title = "{CASA, the Common Astronomy Software Applications for Radio Astronomy}",
      journal = {\pasp},
         year = 2022,
        month = nov,
       volume = {134},
       number = {1041},
          eid = {114501},
        pages = {114501},
          doi = {10.1088/1538-3873/ac9642},
archivePrefix = {arXiv},
       eprint = {2210.02276},
 primaryClass = {astro-ph.IM},
       adsurl = {https://ui.adsabs.harvard.edu/abs/2022PASP..134k4501C}
}

@ARTICLE{2008PASP..120..439C,
       author = {{Cotton}, W.~D.},
        title = "{Obit: A Development Environment for Astronomical Algorithms}",
      journal = {\pasp},
         year = 2008,
        month = apr,
       volume = {120},
       number = {866},
        pages = {439},
          doi = {10.1086/586754},
       adsurl = {https://ui.adsabs.harvard.edu/abs/2008PASP..120..439C}
}

@ARTICLE{2024ApJ...975...34F,
       author = {{Frail}, Dale A. and {Polisensky}, Emil and {Hyman}, Scott D. and {Cotton}, William D. and {Kassim}, Namir E. and {Silverstein}, Michele L. and {Sengar}, Rahul and {Kaplan}, David L. and {Calore}, Francesca and {Berteaud}, Joanna and {Clavel}, Ma{\"\i}ca and {Geyer}, Marisa and {Legodi}, Samuel and {Krishnan}, Vasaant and {Buchner}, Sarah and {Camilo}, Fernando},
        title = "{An Image-based Search for Pulsar Candidates in the MeerKAT Bulge Survey}",
      journal = {\apj},
         year = 2024,
        month = nov,
       volume = {975},
       number = {1},
          eid = {34},
        pages = {34},
          doi = {10.3847/1538-4357/ad74fd},
archivePrefix = {arXiv},
       eprint = {2407.01773},
 primaryClass = {astro-ph.HE},
       adsurl = {https://ui.adsabs.harvard.edu/abs/2024ApJ...975...34F}
}

@ARTICLE{2005AJ....129.1993M,
       author = {{Manchester}, R.~N. and {Hobbs}, G.~B. and {Teoh}, A. and {Hobbs}, M.},
        title = "{The Australia Telescope National Facility Pulsar Catalogue}",
      journal = {\aj},
         year = 2005,
        month = apr,
       volume = {129},
       number = {4},
        pages = {1993-2006},
          doi = {10.1086/428488},
archivePrefix = {arXiv},
       eprint = {astro-ph/0412641},
 primaryClass = {astro-ph},
       adsurl = {https://ui.adsabs.harvard.edu/abs/2005AJ....129.1993M}
}

@INPROCEEDINGS{2016mks..confE...1J,
       author = {{Jonas}, J. and {MeerKAT Team}},
        title = "{The MeerKAT Radio Telescope}",
    booktitle = {MeerKAT Science: On the Pathway to the SKA},
         year = 2016,
        month = jan,
          eid = {1},
        pages = {1},
          doi = {10.22323/1.277.0001},
       adsurl = {https://ui.adsabs.harvard.edu/abs/2016mks..confE...1J}
}

@ARTICLE{2026MNRAS.545f2008L,
       author = {{Lee}, Yu Wing Joshua and {Wang}, Yuanming and {Caleb}, Manisha and {Murphy}, Tara and {An}, Tao and {Das}, Barnali and {Dobie}, Dougal and {Driessen}, Laura N. and {Kaplan}, David L. and {Lenc}, Emil and {Pritchard}, Joshua and {Wadiasingh}, Zorawar and {Xu}, Zhijun},
        title = "{Searching for long-period radio transients in ASKAP EMU data with 10-s imaging}",
      journal = {\mnras},
         year = 2026,
        month = jan,
       volume = {545},
       number = {2},
          eid = {staf2008},
        pages = {staf2008},
          doi = {10.1093/mnras/staf2008},
archivePrefix = {arXiv},
       eprint = {2511.09770},
 primaryClass = {astro-ph.HE},
       adsurl = {https://ui.adsabs.harvard.edu/abs/2026MNRAS.545f2008L}
}

@ARTICLE{2026PASA...43....6M,
       author = {{Murphy}, Tara and {Kaplan}, David L.},
        title = "{The Dawes review 13: A new look at the dynamic radio sky}",
      journal = {\pasa},
         year = 2026,
        month = jan,
       volume = {43},
          eid = {e006},
        pages = {e006},
          doi = {10.1017/pasa.2025.10128},
archivePrefix = {arXiv},
       eprint = {2511.10785},
 primaryClass = {astro-ph.SR},
       adsurl = {https://ui.adsabs.harvard.edu/abs/2026PASA...43....6M}
}

@ARTICLE{2021PASA...38...54M,
       author = {{Murphy}, Tara and {Kaplan}, David L. and {Stewart}, Adam J. and {O'Brien}, Andrew and {Lenc}, Emil and {Pintaldi}, Sergio and {Pritchard}, Joshua and {Dobie}, Dougal and {Fox}, Archibald and {Leung}, James K. and {An}, Tao and {Bell}, Martin E. and {Broderick}, Jess W. and {Chatterjee}, Shami and {Dai}, Shi and {d'Antonio}, Daniele and {Doyle}, Gerry and {Gaensler}, B.~M. and {Heald}, George and {Horesh}, Assaf and {Jones}, Megan L. and {McConnell}, David and {Moss}, Vanessa A. and {Raja}, Wasim and {Ramsay}, Gavin and {Ryder}, Stuart and {Sadler}, Elaine M. and {Sivakoff}, Gregory R. and {Wang}, Yuanming and {Wang}, Ziteng and {Wheatland}, Michael S. and {Whiting}, Matthew and {Allison}, James R. and {Anderson}, C.~S. and {Ball}, Lewis and {Bannister}, K. and {Bock}, D.~C. -J. and {Bolton}, R. and {Bunton}, J.~D. and {Chekkala}, R. and {Chippendale}, A.~P. and {Cooray}, F.~R. and {Gupta}, N. and {Hayman}, D.~B. and {Jeganathan}, K. and {Koribalski}, B. and {Lee-Waddell}, K. and {Mahony}, Elizabeth K. and {Marvil}, J. and {McClure-Griffiths}, N.~M. and {Mirtschin}, P. and {Ng}, A. and {Pearce}, S. and {Phillips}, C. and {Voronkov}, M.~A.},
        title = "{The ASKAP Variables and Slow Transients (VAST) Pilot Survey}",
      journal = {\pasa},
         year = 2021,
        month = oct,
       volume = {38},
          eid = {e054},
        pages = {e054},
          doi = {10.1017/pasa.2021.44},
archivePrefix = {arXiv},
       eprint = {2108.06039},
 primaryClass = {astro-ph.HE},
       adsurl = {https://ui.adsabs.harvard.edu/abs/2021PASA...38...54M}
}

@ARTICLE{2022PASA...39...35H,
       author = {{Hurley-Walker}, N. and {Galvin}, T.~J. and {Duchesne}, S.~W. and {Zhang}, X. and {Morgan}, J. and {Hancock}, P.~J. and {An}, T. and {Franzen}, T.~M.~O. and {Heald}, G. and {Ross}, K. and {Vernstrom}, T. and {Anderson}, G.~E. and {Gaensler}, B.~M. and {Johnston-Hollitt}, M. and {Kaplan}, D.~L. and {Riseley}, C.~J. and {Tingay}, S.~J. and {Walker}, M.},
        title = "{GaLactic and Extragalactic All-sky Murchison Widefield Array survey eXtended (GLEAM-X) I: Survey description and initial data release}",
      journal = {\pasa},
         year = 2022,
        month = aug,
       volume = {39},
          eid = {e035},
        pages = {e035},
          doi = {10.1017/pasa.2022.17},
archivePrefix = {arXiv},
       eprint = {2204.12762},
 primaryClass = {astro-ph.GA},
       adsurl = {https://ui.adsabs.harvard.edu/abs/2022PASA...39...35H}
}

@ARTICLE{2026arXiv260418688R,
       author = {{Rodriguez}, Antonio C. and {El-Badry}, Kareem and {de Ruiter}, Iris and {Rajwade}, Kaustubh and {Berger}, Edo and {Connor}, Liam and {Hurley-Walker}, Natasha},
        title = "{White dwarf + M dwarf Detached Binaries in Long Period Radio Transients: Observed Binary Parameters, Evolution, and Population Constraints}",
      journal = {arXiv e-prints},
         year = 2026,
        month = apr,
          eid = {arXiv:2604.18688},
        pages = {arXiv:2604.18688},
          doi = {10.48550/arXiv.2604.18688},
archivePrefix = {arXiv},
       eprint = {2604.18688},
 primaryClass = {astro-ph.SR},
       adsurl = {https://ui.adsabs.harvard.edu/abs/2026arXiv260418688R}
}

@ARTICLE{2024NatAs...8..230K,
       author = {{Kramer}, Michael and {Liu}, Kuo and {Desvignes}, Gregory and {Karuppusamy}, Ramesh and {Stappers}, Ben W.},
        title = "{Quasi-periodic sub-pulse structure as a unifying feature for radio-emitting neutron stars}",
      journal = {Nature Astronomy},
         year = 2024,
        month = feb,
       volume = {8},
        pages = {230-240},
          doi = {10.1038/s41550-023-02125-3},
archivePrefix = {arXiv},
       eprint = {2311.13762},
 primaryClass = {astro-ph.HE},
       adsurl = {https://ui.adsabs.harvard.edu/abs/2024NatAs...8..230K}
}

@INPROCEEDINGS{2010AAS...21547013G,
       author = {{Gaensler}, Bryan M. and {Landecker}, T.~L. and {Taylor}, A.~R. and {POSSUM Collaboration}},
        title = "{Survey Science with ASKAP: Polarization Sky Survey of the Universe's Magnetism (POSSUM)}",
    booktitle = {American Astronomical Society Meeting Abstracts \#215},
         year = 2010,
       series = {American Astronomical Society Meeting Abstracts},
       volume = {215},
        month = jan,
          eid = {470.13},
        pages = {470.13},
       adsurl = {https://ui.adsabs.harvard.edu/abs/2010AAS...21547013G}
}

@ARTICLE{2013PASA...30....6M,
       author = {{Murphy}, Tara and {Chatterjee}, Shami and {Kaplan}, David L. and {Banyer}, Jay and {Bell}, Martin E. and {Bignall}, Hayley E. and {Bower}, Geoffrey C. and {Cameron}, Robert A. and {Coward}, David M. and {Cordes}, James M. and {Croft}, Steve and {Curran}, James R. and {Djorgovski}, S.~G. and {Farrell}, Sean A. and {Frail}, Dale A. and {Gaensler}, B.~M. and {Galloway}, Duncan K. and {Gendre}, Bruce and {Green}, Anne J. and {Hancock}, Paul J. and {Johnston}, Simon and {Kamble}, Atish and {Law}, Casey J. and {Lazio}, T. Joseph W. and {Lo}, Kitty K. and {Macquart}, Jean-Pierre and {Rea}, Nanda and {Rebbapragada}, Umaa and {Reynolds}, Cormac and {Ryder}, Stuart D. and {Schmidt}, Brian and {Soria}, Roberto and {Stairs}, Ingrid H. and {Tingay}, Steven J. and {Torkelsson}, Ulf and {Wagstaff}, Kiri and {Walker}, Mark and {Wayth}, Randall B. and {Williams}, Peter K.~G.},
        title = "{VAST: An ASKAP Survey for Variables and Slow Transients}",
      journal = {\pasa},
         year = 2013,
        month = feb,
       volume = {30},
          eid = {e006},
        pages = {e006},
          doi = {10.1017/pasa.2012.006},
archivePrefix = {arXiv},
       eprint = {1207.1528},
 primaryClass = {astro-ph.IM},
       adsurl = {https://ui.adsabs.harvard.edu/abs/2013PASA...30....6M}
}

@ARTICLE{2018A&A...611A..66S,
       author = {{Stanway}, E.~R. and {Marsh}, T.~R. and {Chote}, P. and {G{\"a}nsicke}, B.~T. and {Steeghs}, D. and {Wheatley}, P.~J.},
        title = "{VLA radio observations of AR Scorpii}",
      journal = {\aap},
         year = 2018,
        month = mar,
       volume = {611},
          eid = {A66},
        pages = {A66},
          doi = {10.1051/0004-6361/201732380},
archivePrefix = {arXiv},
       eprint = {1801.07258},
 primaryClass = {astro-ph.SR},
       adsurl = {https://ui.adsabs.harvard.edu/abs/2018A&A...611A..66S}
}

@ARTICLE{2025ApJ...988L..29D,
       author = {{Dong}, Fengqiu Adam and {Shin}, Kaitlyn and {Law}, Casey and {Ng}, Mason and {Stairs}, Ingrid and {Bower}, Geoffrey and {Cassity}, Alyssa and {Fonseca}, Emmanuel and {Gaensler}, B.~M. and {Hessels}, Jason W.~T. and {Kaspi}, Victoria M. and {Kharel}, Bikash and {Leung}, Calvin and {Main}, Robert A. and {Masui}, Kiyoshi W. and {McKee}, James W. and {Meyers}, Bradley W. and {Modilim}, Obinna and {Pandhi}, Ayush and {Pearlman}, Aaron B. and {Ransom}, Scott M. and {Scholz}, Paul and {Smith}, Kendrick},
        title = "{CHIME/Fast Radio Burst Discovery of an Unusual Circularly Polarized Long-period Radio Transient with an Accelerating Spin Period}",
      journal = {\apjl},
         year = 2025,
        month = jul,
       volume = {988},
       number = {1},
          eid = {L29},
        pages = {L29},
          doi = {10.3847/2041-8213/adeaab},
archivePrefix = {arXiv},
       eprint = {2507.05139},
 primaryClass = {astro-ph.HE},
       adsurl = {https://ui.adsabs.harvard.edu/abs/2025ApJ...988L..29D}
}

@ARTICLE{2025MNRAS.542..203M,
       author = {{McSweeney}, Samuel J. and {Hurley-Walker}, Natasha and {Horv{\'a}th}, Csan{\'a}d and {Anumarlapudi}, Akash and {Waszewski}, Angie and {Dobie}, Dougal and {Kaplan}, David L. and {Morgan}, John and {Rose}, Kovi and {Wang}, Ziteng},
        title = "{A new long-period radio transient: discovery of pulses repeating every 1.16 h from ASKAP J175534.9{\ensuremath{-}}252749.1}",
      journal = {\mnras},
         year = 2025,
        month = sep,
       volume = {542},
       number = {1},
        pages = {203-214},
          doi = {10.1093/mnras/staf1203},
archivePrefix = {arXiv},
       eprint = {2507.14448},
 primaryClass = {astro-ph.HE},
       adsurl = {https://ui.adsabs.harvard.edu/abs/2025MNRAS.542..203M}
}

@ARTICLE{2019MNRAS.490.4565J,
       author = {{Johnston}, Simon and {Kramer}, Michael},
        title = "{On the beam properties of radio pulsars with interpulse emission}",
      journal = {\mnras},
         year = 2019,
        month = dec,
       volume = {490},
       number = {4},
        pages = {4565-4574},
          doi = {10.1093/mnras/stz2865},
archivePrefix = {arXiv},
       eprint = {1910.04550},
 primaryClass = {astro-ph.HE},
       adsurl = {https://ui.adsabs.harvard.edu/abs/2019MNRAS.490.4565J}
}

@ARTICLE{2008MNRAS.387.1755W,
       author = {{Weltevrede}, Patrick and {Johnston}, Simon},
        title = "{The population of pulsars with interpulses and the implications for beam evolution}",
      journal = {\mnras},
         year = 2008,
        month = jul,
       volume = {387},
       number = {4},
        pages = {1755-1760},
          doi = {10.1111/j.1365-2966.2008.13382.x},
archivePrefix = {arXiv},
       eprint = {0804.4318},
 primaryClass = {astro-ph},
       adsurl = {https://ui.adsabs.harvard.edu/abs/2008MNRAS.387.1755W}
}

@ARTICLE{2008MNRAS.390...87K,
       author = {{Kramer}, Michael and {Johnston}, Simon},
        title = "{High-precision geometry of a double-pole pulsar}",
      journal = {\mnras},
         year = 2008,
        month = oct,
       volume = {390},
       number = {1},
        pages = {87-92},
          doi = {10.1111/j.1365-2966.2008.13780.x},
archivePrefix = {arXiv},
       eprint = {0807.5013},
 primaryClass = {astro-ph},
       adsurl = {https://ui.adsabs.harvard.edu/abs/2008MNRAS.390...87K}
}

@ARTICLE{2025NatAs...9..393L,
       author = {{Lee}, Y.~W.~J. and {Caleb}, M. and {Murphy}, Tara and {Lenc}, E. and {Kaplan}, D.~L. and {Ferrario}, L. and {Wadiasingh}, Z. and {Anumarlapudi}, A. and {Hurley-Walker}, N. and {Karambelkar}, V. and {Ocker}, S.~K. and {McSweeney}, S. and {Qiu}, H. and {Rajwade}, K.~M. and {Zic}, A. and {Bannister}, K.~W. and {Bhat}, N.~D.~R. and {Deller}, A. and {Dobie}, D. and {Driessen}, L.~N. and {Gendreau}, K. and {Glowacki}, M. and {Gupta}, V. and {Jahns-Schindler}, J.~N. and {Jaini}, A. and {James}, C.~W. and {Kasliwal}, M.~M. and {Lower}, M.~E. and {Shannon}, R.~M. and {Uttarkar}, P.~A. and {Wang}, Y. and {Wang}, Z.},
        title = "{The emission of interpulses by a 6.45-h-period coherent radio transient}",
      journal = {Nature Astronomy},
         year = 2025,
        month = mar,
       volume = {9},
        pages = {393-405},
          doi = {10.1038/s41550-024-02452-z},
archivePrefix = {arXiv},
       eprint = {2501.09133},
 primaryClass = {astro-ph.HE},
       adsurl = {https://ui.adsabs.harvard.edu/abs/2025NatAs...9..393L}
}

@ARTICLE{2025ApJ...988..175L,
       author = {{Liu}, Xiaohui and {Xu}, Heng and {Niu}, Jiarui and {Zhang}, Yongkun and {Jiang}, Jinchen and {Zhou}, Dejiang and {Han}, Jinlin and {Zhu}, Weiwei and {Lee}, Kejia and {Li}, Di and {Wang}, Wei-Yang and {Zhang}, Bing and {Chen}, Xuelei and {Luo}, Jia-Wei and {Luo}, Rui and {Niu}, Chenhui and {Qu}, Yuanhong and {Wang}, Bojun and {Wang}, Fayin and {Wang}, Pei and {Wang}, Tiancong and {Wu}, Qin and {Wu}, Ziwei and {Xu}, Jiangwei and {Yang}, Yuan-Pei and {Zhang}, Jun-Shuo},
        title = "{Polarization Position Angle Swing and the Rotating Vector Model of Repeating Fast Radio Bursts}",
      journal = {\apj},
         year = 2025,
        month = aug,
       volume = {988},
       number = {2},
          eid = {175},
        pages = {175},
          doi = {10.3847/1538-4357/ade689},
archivePrefix = {arXiv},
       eprint = {2504.00391},
 primaryClass = {astro-ph.HE},
       adsurl = {https://ui.adsabs.harvard.edu/abs/2025ApJ...988..175L}
}

@ARTICLE{2007A&A...463..647B,
       author = {{Beuermann}, K. and {Euchner}, F. and {Reinsch}, K. and {Jordan}, S. and {G{\"a}nsicke}, B.~T.},
        title = "{Zeeman tomography of magnetic white dwarfs. IV. The complex field structure of the polars EF Eridani, BL Hydri and CP Tucanae}",
      journal = {\aap},
         year = 2007,
        month = feb,
       volume = {463},
       number = {2},
        pages = {647-655},
          doi = {10.1051/0004-6361:20066332},
archivePrefix = {arXiv},
       eprint = {astro-ph/0610804},
 primaryClass = {astro-ph},
       adsurl = {https://ui.adsabs.harvard.edu/abs/2007A&A...463..647B}
}

@ARTICLE{2018CoSka..48..228K,
       author = {{Kawka}, A.},
        title = "{The properties and origin of magnetic fields in white dwarfs}",
      journal = {Contributions of the Astronomical Observatory Skalnate Pleso},
         year = 2018,
        month = jan,
       volume = {48},
       number = {1},
        pages = {228-235},
          doi = {10.48550/arXiv.1801.05602},
archivePrefix = {arXiv},
       eprint = {1801.05602},
 primaryClass = {astro-ph.SR},
       adsurl = {https://ui.adsabs.harvard.edu/abs/2018CoSka..48..228K}
}

@ARTICLE{2011BASI...39..333K,
       author = {{Keane}, E.~F. and {McLaughlin}, M.~A.},
        title = "{Rotating radio transients}",
      journal = {Bulletin of the Astronomical Society of India},
         year = 2011,
        month = sep,
       volume = {39},
       number = {3},
        pages = {333-352},
          doi = {10.48550/arXiv.1109.6896},
archivePrefix = {arXiv},
       eprint = {1109.6896},
 primaryClass = {astro-ph.SR},
       adsurl = {https://ui.adsabs.harvard.edu/abs/2011BASI...39..333K}
}

@ARTICLE{2025AN....34650024M,
       author = {{Ma}, Wen-Qi and {Gao}, Zhi-Fu},
        title = "{The Properties of Fast Radio Bursts: A Short Review}",
      journal = {Astronomische Nachrichten},
         year = 2025,
        month = sep,
       volume = {346},
       number = {7-8},
          eid = {e20250024},
        pages = {e20250024},
          doi = {10.1002/asna.20250024},
       adsurl = {https://ui.adsabs.harvard.edu/abs/2025AN....34650024M}
}

@ARTICLE{2017ARA&A..55..261K,
       author = {{Kaspi}, Victoria M. and {Beloborodov}, Andrei M.},
        title = "{Magnetars}",
      journal = {\araa},
         year = 2017,
        month = aug,
       volume = {55},
       number = {1},
        pages = {261-301},
          doi = {10.1146/annurev-astro-081915-023329},
archivePrefix = {arXiv},
       eprint = {1703.00068},
 primaryClass = {astro-ph.HE},
       adsurl = {https://ui.adsabs.harvard.edu/abs/2017ARA&A..55..261K}
}

@ARTICLE{2024NatAs...8.1159C,
       author = {{Caleb}, M. and {Lenc}, E. and {Kaplan}, D.~L. and {Murphy}, T. and {Men}, Y.~P. and {Shannon}, R.~M. and {Ferrario}, L. and {Rajwade}, K.~M. and {Clarke}, T.~E. and {Giacintucci}, S. and {Hurley-Walker}, N. and {Hyman}, S.~D. and {Lower}, M.~E. and {McSweeney}, Sam and {Ravi}, V. and {Barr}, E.~D. and {Buchner}, S. and {Flynn}, C.~M.~L. and {Hessels}, J.~W.~T. and {Kramer}, M. and {Pritchard}, J. and {Stappers}, B.~W.},
        title = "{An emission-state-switching radio transient with a 54-minute period}",
      journal = {Nature Astronomy},
         year = 2024,
        month = sep,
       volume = {8},
        pages = {1159-1168},
          doi = {10.1038/s41550-024-02277-w},
archivePrefix = {arXiv},
       eprint = {2407.12266},
 primaryClass = {astro-ph.HE},
       adsurl = {https://ui.adsabs.harvard.edu/abs/2024NatAs...8.1159C}
}

@ARTICLE{2022ApJ...932...21K,
       author = {{Kao}, Melodie M. and {Pineda}, J. Sebastian},
        title = "{Radio Emission from Binary Ultracool Dwarf Systems}",
      journal = {\apj},
         year = 2022,
        month = jun,
       volume = {932},
       number = {1},
          eid = {21},
        pages = {21},
          doi = {10.3847/1538-4357/ac660b},
archivePrefix = {arXiv},
       eprint = {2206.01754},
 primaryClass = {astro-ph.SR},
       adsurl = {https://ui.adsabs.harvard.edu/abs/2022ApJ...932...21K}
}

@ARTICLE{2023NatAs...7..931P,
       author = {{Pelisoli}, Ingrid and {Marsh}, T.~R. and {Buckley}, David A.~H. and {Heywood}, I. and {Potter}, Stephen. B. and {Schwope}, Axel and {Brink}, Jaco and {Standke}, Annie and {Woudt}, P.~A. and {Parsons}, S.~G. and {Green}, M.~J. and {Kepler}, S.~O. and {Munday}, James and {Romero}, A.~D. and {Breedt}, E. and {Brown}, A.~J. and {Dhillon}, V.~S. and {Dyer}, M.~J. and {Kerry}, P. and {Littlefair}, S.~P. and {Sahman}, D.~I. and {Wild}, J.~F.},
        title = "{A 5.3-min-period pulsing white dwarf in a binary detected from radio to X-rays}",
      journal = {Nature Astronomy},
         year = 2023,
        month = aug,
       volume = {7},
        pages = {931-942},
          doi = {10.1038/s41550-023-01995-x},
archivePrefix = {arXiv},
       eprint = {2306.09272},
 primaryClass = {astro-ph.SR},
       adsurl = {https://ui.adsabs.harvard.edu/abs/2023NatAs...7..931P}
}

@ARTICLE{2017A&A...601L...7M,
       author = {{Marcote}, B. and {Marsh}, T.~R. and {Stanway}, E.~R. and {Paragi}, Z. and {Blanchard}, J.~M.},
        title = "{Towards the origin of the radio emission in AR Scorpii, the first radio-pulsing white dwarf binary}",
      journal = {\aap},
         year = 2017,
        month = may,
       volume = {601},
          eid = {L7},
        pages = {L7},
          doi = {10.1051/0004-6361/201730948},
archivePrefix = {arXiv},
       eprint = {1705.00600},
 primaryClass = {astro-ph.HE},
       adsurl = {https://ui.adsabs.harvard.edu/abs/2017A&A...601L...7M}
}

@ARTICLE{2015SSRv..191..111F,
       author = {{Ferrario}, Lilia and {de Martino}, Domitilla and {G{\"a}nsicke}, Boris T.},
        title = "{Magnetic White Dwarfs}",
      journal = {\ssr},
         year = 2015,
        month = oct,
       volume = {191},
       number = {1-4},
        pages = {111-169},
          doi = {10.1007/s11214-015-0152-0},
archivePrefix = {arXiv},
       eprint = {1504.08072},
 primaryClass = {astro-ph.SR},
       adsurl = {https://ui.adsabs.harvard.edu/abs/2015SSRv..191..111F}
}

@ARTICLE{2016Natur.537..374M,
       author = {{Marsh}, T.~R. and {G{\"a}nsicke}, B.~T. and {H{\"u}mmerich}, S. and {Hambsch}, F.-J. and {Bernhard}, K. and {Lloyd}, C. and {Breedt}, E. and {Stanway}, E.~R. and {Steeghs}, D.~T. and {Parsons}, S.~G. and {Toloza}, O. and {Schreiber}, M.~R. and {Jonker}, P.~G. and {van Roestel}, J. and {Kupfer}, T. and {Pala}, A.~F. and {Dhillon}, V.~S. and {Hardy}, L.~K. and {Littlefair}, S.~P. and {Aungwerojwit}, A. and {Arjyotha}, S. and {Koester}, D. and {Bochinski}, J.~J. and {Haswell}, C.~A. and {Frank}, P. and {Wheatley}, P.~J.},
        title = "{A radio-pulsing white dwarf binary star}",
      journal = {\nat},
         year = 2016,
        month = sep,
       volume = {537},
       number = {7620},
        pages = {374-377},
          doi = {10.1038/nature18620},
archivePrefix = {arXiv},
       eprint = {1607.08265},
 primaryClass = {astro-ph.SR},
       adsurl = {https://ui.adsabs.harvard.edu/abs/2016Natur.537..374M}
}

@ARTICLE{2018MNRAS.474L..61D,
       author = {{Das}, Barnali and {Chandra}, Poonam and {Wade}, Gregg A.},
        title = "{Discovery of electron cyclotron MASER emission from the magnetic Bp star HD 133880 with the Giant Metrewave Radio Telescope}",
      journal = {\mnras},
         year = 2018,
        month = feb,
       volume = {474},
       number = {1},
        pages = {L61-L65},
          doi = {10.1093/mnrasl/slx193},
archivePrefix = {arXiv},
       eprint = {1711.09836},
 primaryClass = {astro-ph.SR},
       adsurl = {https://ui.adsabs.harvard.edu/abs/2018MNRAS.474L..61D}
}

@ARTICLE{2020AdSpR..66.1226B,
       author = {{Barrett}, Paul and {Dieck}, Christopher and {Beasley}, Anthony J. and {Mason}, Paul A. and {Singh}, Kulinder P.},
        title = "{Radio observations of magnetic cataclysmic variables}",
      journal = {Advances in Space Research},
         year = 2020,
        month = sep,
       volume = {66},
       number = {5},
        pages = {1226-1234},
          doi = {10.1016/j.asr.2020.04.007},
archivePrefix = {arXiv},
       eprint = {2004.11418},
 primaryClass = {astro-ph.SR},
       adsurl = {https://ui.adsabs.harvard.edu/abs/2020AdSpR..66.1226B}
}

@ARTICLE{2019ApJ...871..214V,
       author = {{Villadsen}, Jackie and {Hallinan}, Gregg},
        title = "{Ultra-wideband Detection of 22 Coherent Radio Bursts on M Dwarfs}",
      journal = {\apj},
         year = 2019,
        month = feb,
       volume = {871},
       number = {2},
          eid = {214},
        pages = {214},
          doi = {10.3847/1538-4357/aaf88e},
archivePrefix = {arXiv},
       eprint = {1810.00855},
 primaryClass = {astro-ph.SR},
       adsurl = {https://ui.adsabs.harvard.edu/abs/2019ApJ...871..214V}
}

@ARTICLE{2026ApJ..1004..123R,
       author = {{Ridder}, Margaret E. and {Barrett}, Paul E. and {Heinke}, Craig O. and {Sivakoff}, Gregory R.},
        title = "{A Broadband Search for Coherent Radio Emission in Cataclysmic Variables}",
      journal = {\apj},
         year = 2026,
        month = jun,
       volume = {1004},
       number = {1},
          eid = {123},
        pages = {123},
          doi = {10.3847/1538-4357/ae69d5},
archivePrefix = {arXiv},
       eprint = {2603.03209},
 primaryClass = {astro-ph.HE},
       adsurl = {https://ui.adsabs.harvard.edu/abs/2026ApJ..1004..123R}
}

@ARTICLE{2025NatAs...9..672D,
       author = {{de Ruiter}, I. and {Rajwade}, K.~M. and {Bassa}, C.~G. and {Rowlinson}, A. and {Wijers}, R.~A.~M.~J. and {Kilpatrick}, C.~D. and {Stefansson}, G. and {Callingham}, J.~R. and {Hessels}, J.~W.~T. and {Clarke}, T.~E. and {Peters}, W. and {Wijnands}, R.~A.~D. and {Shimwell}, T.~W. and {ter Veen}, S. and {Morello}, V. and {Zeimann}, G.~R. and {Mahadevan}, S.},
        title = "{Sporadic radio pulses from a white dwarf binary at the orbital period}",
      journal = {Nature Astronomy},
         year = 2025,
        month = may,
       volume = {9},
        pages = {672-684},
          doi = {10.1038/s41550-025-02491-0},
archivePrefix = {arXiv},
       eprint = {2408.11536},
 primaryClass = {astro-ph.HE},
       adsurl = {https://ui.adsabs.harvard.edu/abs/2025NatAs...9..672D}
}

@ARTICLE{2025A&A...695L...8R,
       author = {{Rodriguez}, Antonio C.},
        title = "{Spectroscopic detection of a 2.9-hour orbit in a long-period radio transient}",
      journal = {\aap},
         year = 2025,
        month = mar,
       volume = {695},
          eid = {L8},
        pages = {L8},
          doi = {10.1051/0004-6361/202553684},
archivePrefix = {arXiv},
       eprint = {2501.03315},
 primaryClass = {astro-ph.SR},
       adsurl = {https://ui.adsabs.harvard.edu/abs/2025A&A...695L...8R}
}

@ARTICLE{2022Natur.601..526H,
       author = {{Hurley-Walker}, N. and {Zhang}, X. and {Bahramian}, A. and {McSweeney}, S.~J. and {O'Doherty}, T.~N. and {Hancock}, P.~J. and {Morgan}, J.~S. and {Anderson}, G.~E. and {Heald}, G.~H. and {Galvin}, T.~J.},
        title = "{A radio transient with unusually slow periodic emission}",
      journal = {\nat},
         year = 2022,
        month = jan,
       volume = {601},
       number = {7894},
        pages = {526-530},
          doi = {10.1038/s41586-021-04272-x},
archivePrefix = {arXiv},
       eprint = {2503.08033},
 primaryClass = {astro-ph.HE},
       adsurl = {https://ui.adsabs.harvard.edu/abs/2022Natur.601..526H}
}

@ARTICLE{2023Natur.619..487H,
       author = {{Hurley-Walker}, N. and {Rea}, N. and {McSweeney}, S.~J. and {Meyers}, B.~W. and {Lenc}, E. and {Heywood}, I. and {Hyman}, S.~D. and {Men}, Y.~P. and {Clarke}, T.~E. and {Coti Zelati}, F. and {Price}, D.~C. and {Horv{\'a}th}, C. and {Galvin}, T.~J. and {Anderson}, G.~E. and {Bahramian}, A. and {Barr}, E.~D. and {Bhat}, N.~D.~R. and {Caleb}, M. and {Dall'Ora}, M. and {de Martino}, D. and {Giacintucci}, S. and {Morgan}, J.~S. and {Rajwade}, K.~M. and {Stappers}, B. and {Williams}, A.},
        title = "{A long-period radio transient active for three decades}",
      journal = {\nat},
         year = 2023,
        month = jul,
       volume = {619},
       number = {7970},
        pages = {487-490},
          doi = {10.1038/s41586-023-06202-5},
       adsurl = {https://ui.adsabs.harvard.edu/abs/2023Natur.619..487H}
}

@ARTICLE{2026ApJ..1003...91A,
       author = {{Anumarlapudi}, Akash and {Kaplan}, David L. and {Hurley-Walker}, Natasha and {Ocker}, Stella Koch and {Kelson}, Daniel and {Dobie}, Dougal and {Driessen}, Laura and {Murphy}, Tara and {Pritchard}, Joshua},
        title = "{A Sample of Short-lived Galactic Radio Transients from ASKAP VAST}",
      journal = {\apj},
         year = 2026,
        month = may,
       volume = {1003},
       number = {1},
          eid = {91},
        pages = {91},
          doi = {10.3847/1538-4357/ae6112},
archivePrefix = {arXiv},
       eprint = {2604.11881},
 primaryClass = {astro-ph.HE},
       adsurl = {https://ui.adsabs.harvard.edu/abs/2026ApJ..1003...91A}
}

@ARTICLE{2008ApJ...687..262K,
       author = {{Kaplan}, D.~L. and {Hyman}, S.~D. and {Roy}, S. and {Bandyopadhyay}, R.~M. and {Chakrabarty}, D. and {Kassim}, N.~E. and {Lazio}, T.~J.~W. and {Ray}, P.~S.},
        title = "{A Search for the Near-Infrared Counterpart to GCRT J1745-3009}",
      journal = {\apj},
         year = 2008,
        month = nov,
       volume = {687},
       number = {1},
        pages = {262-271},
          doi = {10.1086/591436},
archivePrefix = {arXiv},
       eprint = {0807.1507},
 primaryClass = {astro-ph},
       adsurl = {https://ui.adsabs.harvard.edu/abs/2008ApJ...687..262K}
}

@ARTICLE{2002ARA&A..40..217G,
       author = {{G{\"u}del}, Manuel},
        title = "{Stellar Radio Astronomy: Probing Stellar Atmospheres from Protostars to Giants}",
      journal = {\araa},
         year = 2002,
        month = jan,
       volume = {40},
        pages = {217-261},
          doi = {10.1146/annurev.astro.40.060401.093806},
archivePrefix = {arXiv},
       eprint = {astro-ph/0206436},
 primaryClass = {astro-ph},
       adsurl = {https://ui.adsabs.harvard.edu/abs/2002ARA&A..40..217G}
}

@ARTICLE{2025Natur.642..583W,
       author = {{Wang}, Ziteng and {Rea}, Nanda and {Bao}, Tong and {Kaplan}, David L. and {Lenc}, Emil and {Wadiasingh}, Zorawar and {Hare}, Jeremy and {Zic}, Andrew and {Anumarlapudi}, Akash and {Bera}, Apurba and {Beniamini}, Paz and {Cooper}, A.~J. and {Clarke}, Tracy E. and {Deller}, Adam T. and {Dawson}, J.~R. and {Glowacki}, Marcin and {Hurley-Walker}, Natasha and {McSweeney}, S.~J. and {Polisensky}, Emil J. and {Peters}, Wendy M. and {Younes}, George and {Bannister}, Keith W. and {Caleb}, Manisha and {Dage}, Kristen C. and {James}, Clancy W. and {Kasliwal}, Mansi M. and {Karambelkar}, Viraj and {Lower}, Marcus E. and {Mori}, Kaya and {Ocker}, Stella Koch and {P{\'e}rez-Torres}, Miguel and {Qiu}, Hao and {Rose}, Kovi and {Shannon}, Ryan M. and {Taub}, Rhianna and {Wang}, Fayin and {Wang}, Yuanming and {Zhao}, Zhenyin and {Bhat}, N.~D. Ramesh and {Dobie}, Dougal and {Driessen}, Laura N. and {Murphy}, Tara and {Jaini}, Akhil and {Deng}, Xinping and {Jahns-Schindler}, Joscha N. and {Lee}, Y.~W. Joshua and {Pritchard}, Joshua and {Tuthill}, John and {Thyagarajan}, Nithyanandan},
        title = "{Detection of X-ray emission from a bright long-period radio transient}",
      journal = {\nat},
         year = 2025,
        month = jun,
       volume = {642},
       number = {8068},
        pages = {583-586},
          doi = {10.1038/s41586-025-09077-w},
archivePrefix = {arXiv},
       eprint = {2411.16606},
 primaryClass = {astro-ph.HE},
       adsurl = {https://ui.adsabs.harvard.edu/abs/2025Natur.642..583W}
}

@ARTICLE{2025ApJ...990L..49D,
       author = {{Dong}, Fengqiu Adam and {Clarke}, Tracy E. and {Curtin}, Alice and {Kumar}, Ajay and {Mckinven}, Ryan and {Shin}, Kaitlyn and {Stairs}, Ingrid and {Brar}, Charanjot and {Burdge}, Kevin and {Chatterjee}, Shami and {Cook}, Amanda M. and {Fonseca}, Emmanuel and {Gaensler}, B.~M. and {Hessels}, Jason W. and {Kaspi}, Victoria M. and {Lazda}, Mattias and {Main}, Robert and {Masui}, Kiyoshi W. and {McKee}, James W. and {Meyers}, Bradley W. and {Pearlman}, Aaron B. and {Ransom}, Scott M. and {Scholz}, Paul and {Smith}, Kendrick M. and {Tan}, Chia Min},
        title = "{CHIME/Fast Radio Burst/Pulsar Discovery of a Nearby Long-period Radio Transient with a Timing Glitch}",
      journal = {\apjl},
         year = 2025,
        month = sep,
       volume = {990},
       number = {2},
          eid = {L49},
        pages = {L49},
          doi = {10.3847/2041-8213/adfa8e},
archivePrefix = {arXiv},
       eprint = {2407.07480},
 primaryClass = {astro-ph.HE},
       adsurl = {https://ui.adsabs.harvard.edu/abs/2025ApJ...990L..49D}
}

@ARTICLE{2005Natur.434...50H,
       author = {{Hyman}, Scott D. and {Lazio}, T. Joseph W. and {Kassim}, Namir E. and {Ray}, Paul S. and {Markwardt}, Craig B. and {Yusef-Zadeh}, Farhad},
        title = "{A powerful bursting radio source towards the Galactic Centre}",
      journal = {\nat},
         year = 2005,
        month = mar,
       volume = {434},
       number = {7029},
        pages = {50-52},
          doi = {10.1038/nature03400},
archivePrefix = {arXiv},
       eprint = {astro-ph/0503052},
 primaryClass = {astro-ph},
       adsurl = {https://ui.adsabs.harvard.edu/abs/2005Natur.434...50H}
}

@ARTICLE{2007ApJ...660L.121H,
       author = {{Hyman}, Scott D. and {Roy}, Subhashis and {Pal}, Sabyasachi and {Lazio}, T. Joseph W. and {Ray}, Paul S. and {Kassim}, Namir E. and {Bhatnagar}, S.},
        title = "{A Faint, Steep-Spectrum Burst from the Radio Transient GCRT J1745-3009}",
      journal = {\apjl},
         year = 2007,
        month = may,
       volume = {660},
       number = {2},
        pages = {L121-L124},
          doi = {10.1086/518245},
archivePrefix = {arXiv},
       eprint = {astro-ph/0701098},
 primaryClass = {astro-ph},
       adsurl = {https://ui.adsabs.harvard.edu/abs/2007ApJ...660L.121H}
}

@ARTICLE{2022MNRAS.516.5972W,
       author = {{Wang}, Ziteng and {Murphy}, Tara and {Kaplan}, David L. and {Bannister}, Keith W. and {Lenc}, Emil and {Leung}, James K. and {O'Brien}, Andrew and {Pintaldi}, Sergio and {Pritchard}, Joshua and {Stewart}, Adam J. and {Zic}, Andrew},
        title = "{A pilot ASKAP survey for radio transients towards the Galactic Centre}",
      journal = {\mnras},
         year = 2022,
        month = nov,
       volume = {516},
       number = {4},
        pages = {5972-5988},
          doi = {10.1093/mnras/stac2542},
archivePrefix = {arXiv},
       eprint = {2209.02352},
 primaryClass = {astro-ph.HE},
       adsurl = {https://ui.adsabs.harvard.edu/abs/2022MNRAS.516.5972W}
}

@ARTICLE{2025PASA...42...38D,
       author = {{Duchesne}, S. and {Ross}, K. and {Thomson}, A.~J.~M. and {Lenc}, E. and {Murphy}, T. and {Galvin}, T.~J. and {Hotan}, A.~W. and {Moss}, V.~A. and {Whiting}, M.~T.},
        title = "{The Rapid ASKAP Continuum Survey (RACS) VI: The RACS-high 1655.5 MHz images and catalogue.}",
      journal = {\pasa},
         year = 2025,
        month = jan,
       volume = {42},
          eid = {38},
        pages = {38},
          doi = {10.1017/pasa.2025.2},
archivePrefix = {arXiv},
       eprint = {2501.04978},
 primaryClass = {astro-ph.GA},
       adsurl = {https://ui.adsabs.harvard.edu/abs/2025PASA...42...38D}
}

@ARTICLE{2024PASA...41....3D,
       author = {{Duchesne}, S.~W. and {Grundy}, J.~A. and {Heald}, George H. and {Lenc}, Emil and {Leung}, James K. and {McConnell}, David and {Murphy}, Tara and {Pritchard}, Joshua and {Rose}, Kovi and {Thomson}, Alec J.~M. and {Wang}, Yuanming and {Wang}, Ziteng and {Whiting}, Matthew T.},
        title = "{The Rapid ASKAP Continuum Survey V: Cataloguing the sky at 1367.5 MHz and the second data release of RACS-mid}",
      journal = {\pasa},
         year = 2024,
        month = jan,
       volume = {41},
          eid = {e003},
        pages = {e003},
          doi = {10.1017/pasa.2023.60},
archivePrefix = {arXiv},
       eprint = {2311.12369},
 primaryClass = {astro-ph.GA},
       adsurl = {https://ui.adsabs.harvard.edu/abs/2024PASA...41....3D}
}

@ARTICLE{2016ApJ...832...60P,
       author = {{Polisensky}, E. and {Lane}, W.~M. and {Hyman}, S.~D. and {Kassim}, N.~E. and {Giacintucci}, S. and {Clarke}, T.~E. and {Cotton}, W.~D. and {Cleland}, E. and {Frail}, D.~A.},
        title = "{Exploring the Transient Radio Sky with VLITE: Early Results}",
      journal = {\apj},
         year = 2016,
        month = nov,
       volume = {832},
       number = {1},
          eid = {60},
        pages = {60},
          doi = {10.3847/0004-637X/832/1/60},
archivePrefix = {arXiv},
       eprint = {1604.00667},
 primaryClass = {astro-ph.HE},
       adsurl = {https://ui.adsabs.harvard.edu/abs/2016ApJ...832...60P}
}

@ARTICLE{2024ApJ...976L..21H,
       author = {{Hurley-Walker}, N. and {McSweeney}, S.~J. and {Bahramian}, A. and {Rea}, N. and {Horv{\'a}th}, C. and {Buchner}, S. and {Williams}, A. and {Meyers}, B.~W. and {Strader}, Jay and {Aydi}, Elias and {Urquhart}, Ryan and {Chomiuk}, Laura and {Galvin}, T.~J. and {Coti Zelati}, F. and {Bailes}, Matthew},
        title = "{A 2.9 hr Periodic Radio Transient with an Optical Counterpart}",
      journal = {\apjl},
         year = 2024,
        month = dec,
       volume = {976},
       number = {2},
          eid = {L21},
        pages = {L21},
          doi = {10.3847/2041-8213/ad890e},
archivePrefix = {arXiv},
       eprint = {2408.15757},
 primaryClass = {astro-ph.SR},
       adsurl = {https://ui.adsabs.harvard.edu/abs/2024ApJ...976L..21H}
}

@ARTICLE{2026arXiv260307857P,
       author = {{Pritchard}, Joshua and {Murphy}, Tara and {Dobie}, Dougal and {Lenc}, Emil and {Anumarlapudi}, Akash and {Caleb}, Manisha and {Grainger}, Sophia and {Hurley-Walker}, Natasha and {Kaplan}, David L. and {McSweeney}, Samuel J. and {Mitchell-Bolton}, Jackson and {Rose}, Kovi and {Sengar}, Rahul and {Wang}, Ziteng and {Willingham}, Jayde and {Zic}, Andrew},
        title = "{Discovery of a 36-minute long-period transient ASKAP J142431.2-612611}",
      journal = {arXiv e-prints},
         year = 2026,
        month = mar,
          eid = {arXiv:2603.07857},
        pages = {arXiv:2603.07857},
          doi = {10.48550/arXiv.2603.07857},
archivePrefix = {arXiv},
       eprint = {2603.07857},
 primaryClass = {astro-ph.HE},
       adsurl = {https://ui.adsabs.harvard.edu/abs/2026arXiv260307857P}
}

@ARTICLE{2024arXiv240303127D,
       author = {{Doroshenko}, Victor},
        title = "{3D-$N_{\rm H}$-tool}",
      journal = {arXiv e-prints},
         year = 2024,
        month = mar,
          eid = {arXiv:2403.03127},
        pages = {arXiv:2403.03127},
          doi = {10.48550/arXiv.2403.03127},
archivePrefix = {arXiv},
       eprint = {2403.03127},
 primaryClass = {astro-ph.HE},
       adsurl = {https://ui.adsabs.harvard.edu/abs/2024arXiv240303127D}
}

@ARTICLE{2013PASA...30....7T,
       author = {{Tingay}, S.~J. and {Goeke}, R. and {Bowman}, J.~D. and {Emrich}, D. and {Ord}, S.~M. and {Mitchell}, D.~A. and {Morales}, M.~F. and {Booler}, T. and {Crosse}, B. and {Wayth}, R.~B. and {Lonsdale}, C.~J. and {Tremblay}, S. and {Pallot}, D. and {Colegate}, T. and {Wicenec}, A. and {Kudryavtseva}, N. and {Arcus}, W. and {Barnes}, D. and {Bernardi}, G. and {Briggs}, F. and {Burns}, S. and {Bunton}, J.~D. and {Cappallo}, R.~J. and {Corey}, B.~E. and {Deshpande}, A. and {Desouza}, L. and {Gaensler}, B.~M. and {Greenhill}, L.~J. and {Hall}, P.~J. and {Hazelton}, B.~J. and {Herne}, D. and {Hewitt}, J.~N. and {Johnston-Hollitt}, M. and {Kaplan}, D.~L. and {Kasper}, J.~C. and {Kincaid}, B.~B. and {Koenig}, R. and {Kratzenberg}, E. and {Lynch}, M.~J. and {Mckinley}, B. and {Mcwhirter}, S.~R. and {Morgan}, E. and {Oberoi}, D. and {Pathikulangara}, J. and {Prabu}, T. and {Remillard}, R.~A. and {Rogers}, A.~E.~E. and {Roshi}, A. and {Salah}, J.~E. and {Sault}, R.~J. and {Udaya-Shankar}, N. and {Schlagenhaufer}, F. and {Srivani}, K.~S. and {Stevens}, J. and {Subrahmanyan}, R. and {Waterson}, M. and {Webster}, R.~L. and {Whitney}, A.~R. and {Williams}, A. and {Williams}, C.~L. and {Wyithe}, J.~S.~B.},
        title = "{The Murchison Widefield Array: The Square Kilometre Array Precursor at Low Radio Frequencies}",
      journal = {\pasa},
         year = 2013,
        month = jan,
       volume = {30},
          eid = {e007},
        pages = {e007},
          doi = {10.1017/pasa.2012.007},
archivePrefix = {arXiv},
       eprint = {1206.6945},
 primaryClass = {astro-ph.IM},
       adsurl = {https://ui.adsabs.harvard.edu/abs/2013PASA...30....7T}
}

@ARTICLE{2018PASA...35...33W,
       author = {{Wayth}, Randall B. and {Tingay}, Steven J. and {Trott}, Cathryn M. and {Emrich}, David and {Johnston-Hollitt}, Melanie and {McKinley}, Ben and {Gaensler}, B.~M. and {Beardsley}, A.~P. and {Booler}, T. and {Crosse}, B. and {Franzen}, T.~M.~O. and {Horsley}, L. and {Kaplan}, D.~L. and {Kenney}, D. and {Morales}, M.~F. and {Pallot}, D. and {Sleap}, G. and {Steele}, K. and {Walker}, M. and {Williams}, A. and {Wu}, C. and {Cairns}, Iver. H. and {Filipovic}, M.~D. and {Johnston}, S. and {Murphy}, T. and {Quinn}, P. and {Staveley-Smith}, L. and {Webster}, R. and {Wyithe}, J.~S.~B.},
        title = "{The Phase II Murchison Widefield Array: Design overview}",
      journal = {\pasa},
         year = 2018,
        month = nov,
       volume = {35},
          eid = {e033},
        pages = {e033},
          doi = {10.1017/pasa.2018.37},
archivePrefix = {arXiv},
       eprint = {1809.06466},
 primaryClass = {astro-ph.IM},
       adsurl = {https://ui.adsabs.harvard.edu/abs/2018PASA...35...33W}
}

@ARTICLE{2025MNRAS.542.1208A,
       author = {{Anumarlapudi}, Akash and {Kaplan}, David L. and {Rea}, Nanda and {Erasmus}, Nicolas and {Kelson}, Daniel and {Ocker}, Stella Koch and {Lenc}, Emil and {Dobie}, Dougal and {Hurley-Walker}, Natasha and {Sivakoff}, Gregory and {Buckley}, David A.~H. and {Murphy}, Tara and {Pritchard}, Joshua and {Driessen}, Laura and {Rose}, Kovi and {Zic}, Andrew},
        title = "{ASKAP J144834{\ensuremath{-}}685644: a newly discovered long period radio transient detected from radio to X-rays}",
      journal = {\mnras},
         year = 2025,
        month = sep,
       volume = {542},
       number = {2},
        pages = {1208-1232},
          doi = {10.1093/mnras/staf1227},
archivePrefix = {arXiv},
       eprint = {2507.13453},
 primaryClass = {astro-ph.HE},
       adsurl = {https://ui.adsabs.harvard.edu/abs/2025MNRAS.542.1208A}
}

@ARTICLE{2023A&A...674A...1G,
       author = {{Gaia Collaboration} and {Vallenari}, A. and {Brown}, A.~G.~A. and {Prusti}, T. and {de Bruijne}, J.~H.~J. and {Arenou}, F. and {Babusiaux}, C. and {Biermann}, M. and {Creevey}, O.~L. and {Ducourant}, C. and {Evans}, D.~W. and {Eyer}, L. and {Guerra}, R. and {Hutton}, A. and {Jordi}, C. and {Klioner}, S.~A. and {Lammers}, U.~L. and {Lindegren}, L. and {Luri}, X. and {Mignard}, F. and {Panem}, C. and {Pourbaix}, D. and {Randich}, S. and {Sartoretti}, P. and {Soubiran}, C. and {Tanga}, P. and {Walton}, N.~A. and {Bailer-Jones}, C.~A.~L. and {Bastian}, U. and {Drimmel}, R. and {Jansen}, F. and {Katz}, D. and {Lattanzi}, M.~G. and {van Leeuwen}, F. and {Bakker}, J. and {Cacciari}, C. and {Casta{\~n}eda}, J. and {De Angeli}, F. and {Fabricius}, C. and {Fouesneau}, M. and {Fr{\'e}mat}, Y. and {Galluccio}, L. and {Guerrier}, A. and {Heiter}, U. and {Masana}, E. and {Messineo}, R. and {Mowlavi}, N. and {Nicolas}, C. and {Nienartowicz}, K. and {Pailler}, F. and {Panuzzo}, P. and {Riclet}, F. and {Roux}, W. and {Seabroke}, G.~M. and {Sordo}, R. and {Th{\'e}venin}, F. and {Gracia-Abril}, G. and {Portell}, J. and {Teyssier}, D. and {Altmann}, M. and {Andrae}, R. and {Audard}, M. and {Bellas-Velidis}, I. and {Benson}, K. and {Berthier}, J. and {Blomme}, R. and {Burgess}, P.~W. and {Busonero}, D. and {Busso}, G. and {C{\'a}novas}, H. and {Carry}, B. and {Cellino}, A. and {Cheek}, N. and {Clementini}, G. and {Damerdji}, Y. and {Davidson}, M. and {de Teodoro}, P. and {Nu{\~n}ez Campos}, M. and {Delchambre}, L. and {Dell'Oro}, A. and {Esquej}, P. and {Fern{\'a}ndez-Hern{\'a}ndez}, J. and {Fraile}, E. and {Garabato}, D. and {Garc{\'\i}a-Lario}, P. and {Gosset}, E. and {Haigron}, R. and {Halbwachs}, J. -L. and {Hambly}, N.~C. and {Harrison}, D.~L. and {Hern{\'a}ndez}, J. and {Hestroffer}, D. and {Hodgkin}, S.~T. and {Holl}, B. and {Jan{\ss}en}, K. and {Jevardat de Fombelle}, G. and {Jordan}, S. and {Krone-Martins}, A. and {Lanzafame}, A.~C. and {L{\"o}ffler}, W. and {Marchal}, O. and {Marrese}, P.~M. and {Moitinho}, A. and {Muinonen}, K. and {Osborne}, P. and {Pancino}, E. and {Pauwels}, T. and {Recio-Blanco}, A. and {Reyl{\'e}}, C. and {Riello}, M. and {Rimoldini}, L. and {Roegiers}, T. and {Rybizki}, J. and {Sarro}, L.~M. and {Siopis}, C. and {Smith}, M. and {Sozzetti}, A. and {Utrilla}, E. and {van Leeuwen}, M. and {Abbas}, U. and {{\'A}brah{\'a}m}, P. and {Abreu Aramburu}, A. and {Aerts}, C. and {Aguado}, J.~J. and {Ajaj}, M. and {Aldea-Montero}, F. and {Altavilla}, G. and {{\'A}lvarez}, M.~A. and {Alves}, J. and {Anders}, F. and {Anderson}, R.~I. and {Anglada Varela}, E. and {Antoja}, T. and {Baines}, D. and {Baker}, S.~G. and {Balaguer-N{\'u}{\~n}ez}, L. and {Balbinot}, E. and {Balog}, Z. and {Barache}, C. and {Barbato}, D. and {Barros}, M. and {Barstow}, M.~A. and {Bartolom{\'e}}, S. and {Bassilana}, J. -L. and {Bauchet}, N. and {Becciani}, U. and {Bellazzini}, M. and {Berihuete}, A. and {Bernet}, M. and {Bertone}, S. and {Bianchi}, L. and {Binnenfeld}, A. and {Blanco-Cuaresma}, S. and {Blazere}, A. and {Boch}, T. and {Bombrun}, A. and {Bossini}, D. and {Bouquillon}, S. and {Bragaglia}, A. and {Bramante}, L. and {Breedt}, E. and {Bressan}, A. and {Brouillet}, N. and {Brugaletta}, E. and {Bucciarelli}, B. and {Burlacu}, A. and {Butkevich}, A.~G. and {Buzzi}, R. and {Caffau}, E. and {Cancelliere}, R. and {Cantat-Gaudin}, T. and {Carballo}, R. and {Carlucci}, T. and {Carnerero}, M.~I. and {Carrasco}, J.~M. and {Casamiquela}, L. and {Castellani}, M. and {Castro-Ginard}, A. and {Chaoul}, L. and {Charlot}, P. and {Chemin}, L. and {Chiaramida}, V. and {Chiavassa}, A. and {Chornay}, N. and {Comoretto}, G. and {Contursi}, G. and {Cooper}, W.~J. and {Cornez}, T. and {Cowell}, S. and {Crifo}, F. and {Cropper}, M. and {Crosta}, M. and {Crowley}, C. and {Dafonte}, C. and {Dapergolas}, A. and {David}, M. and {David}, P. and {de Laverny}, P. and {De Luise}, F. and {De March}, R. and {De Ridder}, J. and {de Souza}, R. and {de Torres}, A. and {del Peloso}, E.~F. and {del Pozo}, E. and {Delbo}, M. and {Delgado}, A. and {Delisle}, J. -B. and {Demouchy}, C. and {Dharmawardena}, T.~E. and {Di Matteo}, P. and {Diakite}, S. and {Diener}, C. and {Distefano}, E. and {Dolding}, C. and {Edvardsson}, B. and {Enke}, H. and {Fabre}, C. and {Fabrizio}, M. and {Faigler}, S. and {Fedorets}, G. and {Fernique}, P. and {Fienga}, A. and {Figueras}, F. and {Fournier}, Y. and {Fouron}, C. and {Fragkoudi}, F. and {Gai}, M. and {Garcia-Gutierrez}, A. and {Garcia-Reinaldos}, M. and {Garc{\'\i}a-Torres}, M. and {Garofalo}, A. and {Gavel}, A. and {Gavras}, P. and {Gerlach}, E. and {Geyer}, R. and {Giacobbe}, P. and {Gilmore}, G. and {Girona}, S. and {Giuffrida}, G. and {Gomel}, R. and {Gomez}, A. and {Gonz{\'a}lez-N{\'u}{\~n}ez}, J. and {Gonz{\'a}lez-Santamar{\'\i}a}, I. and {Gonz{\'a}lez-Vidal}, J.~J. and {Granvik}, M. and {Guillout}, P. and {Guiraud}, J. and {Guti{\'e}rrez-S{\'a}nchez}, R. and {Guy}, L.~P. and {Hatzidimitriou}, D. and {Hauser}, M. and {Haywood}, M. and {Helmer}, A. and {Helmi}, A. and {Sarmiento}, M.~H. and {Hidalgo}, S.~L. and {Hilger}, T. and {H{\l}adczuk}, N. and {Hobbs}, D. and {Holland}, G. and {Huckle}, H.~E. and {Jardine}, K. and {Jasniewicz}, G. and {Jean-Antoine Piccolo}, A. and {Jim{\'e}nez-Arranz}, {\'O}. and {Jorissen}, A. and {Juaristi Campillo}, J. and {Julbe}, F. and {Karbevska}, L. and {Kervella}, P. and {Khanna}, S. and {Kontizas}, M. and {Kordopatis}, G. and {Korn}, A.~J. and {K{\'o}sp{\'a}l}, {\'A}. and {Kostrzewa-Rutkowska}, Z. and {Kruszy{\'n}ska}, K. and {Kun}, M. and {Laizeau}, P. and {Lambert}, S. and {Lanza}, A.~F. and {Lasne}, Y. and {Le Campion}, J. -F. and {Lebreton}, Y. and {Lebzelter}, T. and {Leccia}, S. and {Leclerc}, N. and {Lecoeur-Taibi}, I. and {Liao}, S. and {Licata}, E.~L. and {Lindstr{\o}m}, H.~E.~P. and {Lister}, T.~A. and {Livanou}, E. and {Lobel}, A. and {Lorca}, A. and {Loup}, C. and {Madrero Pardo}, P. and {Magdaleno Romeo}, A. and {Managau}, S. and {Mann}, R.~G. and {Manteiga}, M. and {Marchant}, J.~M. and {Marconi}, M. and {Marcos}, J. and {Marcos Santos}, M.~M.~S. and {Mar{\'\i}n Pina}, D. and {Marinoni}, S. and {Marocco}, F. and {Marshall}, D.~J. and {Martin Polo}, L. and {Mart{\'\i}n-Fleitas}, J.~M. and {Marton}, G. and {Mary}, N. and {Masip}, A. and {Massari}, D. and {Mastrobuono-Battisti}, A. and {Mazeh}, T. and {McMillan}, P.~J. and {Messina}, S. and {Michalik}, D. and {Millar}, N.~R. and {Mints}, A. and {Molina}, D. and {Molinaro}, R. and {Moln{\'a}r}, L. and {Monari}, G. and {Mongui{\'o}}, M. and {Montegriffo}, P. and {Montero}, A. and {Mor}, R. and {Mora}, A. and {Morbidelli}, R. and {Morel}, T. and {Morris}, D. and {Muraveva}, T. and {Murphy}, C.~P. and {Musella}, I. and {Nagy}, Z. and {Noval}, L. and {Oca{\~n}a}, F. and {Ogden}, A. and {Ordenovic}, C. and {Osinde}, J.~O. and {Pagani}, C. and {Pagano}, I. and {Palaversa}, L. and {Palicio}, P.~A. and {Pallas-Quintela}, L. and {Panahi}, A. and {Payne-Wardenaar}, S. and {Pe{\~n}alosa Esteller}, X. and {Penttil{\"a}}, A. and {Pichon}, B. and {Piersimoni}, A.~M. and {Pineau}, F. -X. and {Plachy}, E. and {Plum}, G. and {Poggio}, E. and {Pr{\v{s}}a}, A. and {Pulone}, L. and {Racero}, E. and {Ragaini}, S. and {Rainer}, M. and {Raiteri}, C.~M. and {Rambaux}, N. and {Ramos}, P. and {Ramos-Lerate}, M. and {Re Fiorentin}, P. and {Regibo}, S. and {Richards}, P.~J. and {Rios Diaz}, C. and {Ripepi}, V. and {Riva}, A. and {Rix}, H. -W. and {Rixon}, G. and {Robichon}, N. and {Robin}, A.~C. and {Robin}, C. and {Roelens}, M. and {Rogues}, H.~R.~O. and {Rohrbasser}, L. and {Romero-G{\'o}mez}, M. and {Rowell}, N. and {Royer}, F. and {Ruz Mieres}, D. and {Rybicki}, K.~A. and {Sadowski}, G. and {S{\'a}ez N{\'u}{\~n}ez}, A. and {Sagrist{\`a} Sell{\'e}s}, A. and {Sahlmann}, J. and {Salguero}, E. and {Samaras}, N. and {Sanchez Gimenez}, V. and {Sanna}, N. and {Santove{\~n}a}, R. and {Sarasso}, M. and {Schultheis}, M. and {Sciacca}, E. and {Segol}, M. and {Segovia}, J.~C. and {S{\'e}gransan}, D. and {Semeux}, D. and {Shahaf}, S. and {Siddiqui}, H.~I. and {Siebert}, A. and {Siltala}, L. and {Silvelo}, A. and {Slezak}, E. and {Slezak}, I. and {Smart}, R.~L. and {Snaith}, O.~N. and {Solano}, E. and {Solitro}, F. and {Souami}, D. and {Souchay}, J. and {Spagna}, A. and {Spina}, L. and {Spoto}, F. and {Steele}, I.~A. and {Steidelm{\"u}ller}, H. and {Stephenson}, C.~A. and {S{\"u}veges}, M. and {Surdej}, J. and {Szabados}, L. and {Szegedi-Elek}, E. and {Taris}, F. and {Taylor}, M.~B. and {Teixeira}, R. and {Tolomei}, L. and {Tonello}, N. and {Torra}, F. and {Torra}, J. and {Torralba Elipe}, G. and {Trabucchi}, M. and {Tsounis}, A.~T. and {Turon}, C. and {Ulla}, A. and {Unger}, N. and {Vaillant}, M.~V. and {van Dillen}, E. and {van Reeven}, W. and {Vanel}, O. and {Vecchiato}, A. and {Viala}, Y. and {Vicente}, D. and {Voutsinas}, S. and {Weiler}, M. and {Wevers}, T. and {Wyrzykowski}, {\L}. and {Yoldas}, A. and {Yvard}, P. and {Zhao}, H. and {Zorec}, J. and {Zucker}, S. and {Zwitter}, T.},
        title = "{Gaia Data Release 3. Summary of the content and survey properties}",
      journal = {\aap},
         year = 2023,
        month = jun,
       volume = {674},
          eid = {A1},
        pages = {A1},
          doi = {10.1051/0004-6361/202243940},
archivePrefix = {arXiv},
       eprint = {2208.00211},
 primaryClass = {astro-ph.GA},
       adsurl = {https://ui.adsabs.harvard.edu/abs/2023A&A...674A...1G}
}

@ARTICLE{2010ApJS..189...37E,
       author = {{Evans}, Ian N. and {Primini}, Francis A. and {Glotfelty}, Kenny J. and {Anderson}, Craig S. and {Bonaventura}, Nina R. and {Chen}, Judy C. and {Davis}, John E. and {Doe}, Stephen M. and {Evans}, Janet D. and {Fabbiano}, Giuseppina and {Galle}, Elizabeth C. and {Gibbs}, Danny G., II and {Grier}, John D. and {Hain}, Roger M. and {Hall}, Diane M. and {Harbo}, Peter N. and {He}, Xiangqun Helen and {Houck}, John C. and {Karovska}, Margarita and {Kashyap}, Vinay L. and {Lauer}, Jennifer and {McCollough}, Michael L. and {McDowell}, Jonathan C. and {Miller}, Joseph B. and {Mitschang}, Arik W. and {Morgan}, Douglas L. and {Mossman}, Amy E. and {Nichols}, Joy S. and {Nowak}, Michael A. and {Plummer}, David A. and {Refsdal}, Brian L. and {Rots}, Arnold H. and {Siemiginowska}, Aneta and {Sundheim}, Beth A. and {Tibbetts}, Michael S. and {Van Stone}, David W. and {Winkelman}, Sherry L. and {Zografou}, Panagoula},
        title = "{The Chandra Source Catalog}",
      journal = {\apjs},
         year = 2010,
        month = jul,
       volume = {189},
       number = {1},
        pages = {37-82},
          doi = {10.1088/0067-0049/189/1/37},
archivePrefix = {arXiv},
       eprint = {1005.4665},
 primaryClass = {astro-ph.HE},
       adsurl = {https://ui.adsabs.harvard.edu/abs/2010ApJS..189...37E}
}

@ARTICLE{2010AJ....140.1868W,
       author = {{Wright}, Edward L. and {Eisenhardt}, Peter R.~M. and {Mainzer}, Amy K. and {Ressler}, Michael E. and {Cutri}, Roc M. and {Jarrett}, Thomas and {Kirkpatrick}, J. Davy and {Padgett}, Deborah and {McMillan}, Robert S. and {Skrutskie}, Michael and {Stanford}, S.~A. and {Cohen}, Martin and {Walker}, Russell G. and {Mather}, John C. and {Leisawitz}, David and {Gautier}, Thomas N., III and {McLean}, Ian and {Benford}, Dominic and {Lonsdale}, Carol J. and {Blain}, Andrew and {Mendez}, Bryan and {Irace}, William R. and {Duval}, Valerie and {Liu}, Fengchuan and {Royer}, Don and {Heinrichsen}, Ingolf and {Howard}, Joan and {Shannon}, Mark and {Kendall}, Martha and {Walsh}, Amy L. and {Larsen}, Mark and {Cardon}, Joel G. and {Schick}, Scott and {Schwalm}, Mark and {Abid}, Mohamed and {Fabinsky}, Beth and {Naes}, Larry and {Tsai}, Chao-Wei},
        title = "{The Wide-field Infrared Survey Explorer (WISE): Mission Description and Initial On-orbit Performance}",
      journal = {\aj},
         year = 2010,
        month = dec,
       volume = {140},
       number = {6},
        pages = {1868-1881},
          doi = {10.1088/0004-6256/140/6/1868},
archivePrefix = {arXiv},
       eprint = {1008.0031},
 primaryClass = {astro-ph.IM},
       adsurl = {https://ui.adsabs.harvard.edu/abs/2010AJ....140.1868W}
}

@MISC{2015ascl.soft02007M,
       author = {{Mohan}, Niruj and {Rafferty}, David},
        title = "{PyBDSF: Python Blob Detection and Source Finder}",
 howpublished = {Astrophysics Source Code Library, record ascl:1502.007},
         year = 2015,
        month = feb,
          eid = {ascl:1502.007},
        pages = {ascl:1502.007},
archivePrefix = {ascl},
       eprint = {1502.007},
       adsurl = {https://ui.adsabs.harvard.edu/abs/2015ascl.soft02007M}
}

@misc{https://doi.org/10.48479/nz0n-p845,
doi = {10.48479/NZ0N-P845},
url = {https://archive-gw-1.kat.ac.za/public/repository/10.48479/nz0n-p845/index.html},
author = {Cotton, W. D. and Kothes, R. and Camilo, F. and Chandra, P. and Bucher, S. and Nyamai, M.},
title = {MeerKAT 1.3 GHz Observations of Supernove Remnants},
publisher = {South African Radio Astronomy Observatory},
year = {2024}
}

@misc{Thomson2026AlecThomson,
author = {Thomson, Alec and Duchesne, Stef and {Galvin}, Timothy J.},
year = {2026},
month = {jul 13},
title = {AlecThomson/{FixMS}},
url = {https://github.com/AlecThomson/FixMS},
publisher = {GitHub},
howpublished = {https://github.com/AlecThomson/FixMS},
}

@misc{https://doi.org/10.25919/5cca3787a6353,
doi = {10.25919/5CCA3787A6353},
url = {https://data.csiro.au/collections/#collection/CIcsiro:39526v1/DItrue},
author = {Guzman, Juan and Whiting, Matthew and Voronkov, Max and Mitchell, Daniel and Ord, Stephen and Collins, Daniel and Marquarding, Malte and Lahur, Paulus and Maher, Tony and Van Diepen, Ger and Bannister, Keith and Wu, Xinyu and Lenc, Emil and Khoo, Jonathan and Bastholm, Eric},
title = {ASKAP Science Data Processor software - ASKAPsoft Version 0.23.3},
publisher = {CSIRO},
year = {2019}
}
\bibliographystyle{aasjournal}

\end{document}